\documentclass[final,5p,letterpaper,times,twocolumn]{elsarticle}

\usepackage{amssymb}

\usepackage{amssymb}
\usepackage{graphicx} 
\usepackage{multirow}
\usepackage{booktabs} 
\usepackage{bm}
\usepackage{array} 
\usepackage{appendix}
\usepackage{epstopdf}
\usepackage{bbm}
\usepackage{algorithm}
\usepackage{algorithmicx}
\usepackage{algpseudocode}
\usepackage{hyperref}

\usepackage{amsmath,dsfont}
\usepackage{cases,cleveref}
\crefname{equation}{Eq.}{Eqs.}
\usepackage{harpoon}
\usepackage{amssymb}
\usepackage{mathrsfs}
\usepackage{graphics}
\usepackage{color}
\usepackage{geometry}
\usepackage{epsfig}
\usepackage{diagbox}
\usepackage{mathtools}
\usepackage[final]{changes}
\usepackage{comment}
\usepackage{bbm}
\usepackage{lineno}
\usepackage{cuted}
\usepackage{graphicx}
\def\c{{\bf c}}

\def\dd{\mathrm{d}}

\def\g{\gamma}

\def\q{{\bf q}}

\def\x{{\bf x}}

\def\l{\ell}

\usepackage{wrapfig}
\usepackage{dsfont}
\usepackage{enumerate}
\usepackage{subfigure}
\usepackage{xcolor}
\usepackage{setspace}

\usepackage{enumitem}
\setitemize{leftmargin=2cm}

\journal{Mathematical Biosciences}

\begin{document}

\begin{frontmatter}
	
\title{A probabilistic model of relapse in drug addiction}

\author[1]{Sayun Mao}
\affiliation[1]{organization={Department of Computational Medicine},
  addressline={UCLA},
  city={Los Angeles},
  postcode={90095-1766},
  state={CA}, country={USA}}
\ead{maosayun@ucla.edu}

\author[2]{Tom Chou}
\affiliation[2]{organization={Department of Computational Medicine},
  addressline={UCLA},
  city={Los Angeles},
  postcode={90095-1766},
  state={CA}, country={USA}}
\ead{tomchou@ucla.edu}

\author[2,3]{Maria R. D'Orsogna \corref{cor1}}
\ead{dorsogna@csun.edu}
\cortext[cor1]{Corresponding author}
\affiliation[3]{organization={Department of Mathematics},
  addressline={California State University at Northridge},
  city={Los Angeles},
  postcode={91330},
  state={CA}, country={USA}}

\begin{abstract}
More than 60$\%$ of individuals recovering from substance use disorder
relapse within one year. Some will resume drug consumption even after
decades of abstinence. The cognitive and psychological mechanisms that
lead to relapse are not completely understood, but stressful life
experiences and external stimuli that are associated with past
drug-taking are known to play a primary role.  Stressors and cues
elicit memories of drug-induced euphoria and the expectation of relief
from current anxiety, igniting an intense craving to use again;
positive experiences and supportive environments may mitigate relapse.
We present a mathematical model of relapse in drug addiction that
draws on known psychiatric concepts such as the ``positive activation;
negative activation'' paradigm and the ``peak-end'' rule to construct
a relapse rate that depends on external factors (intensity and timing
of life events) and individual traits (mental responses to these
events).  We analyze which combinations and ordering of stressors,
cues, and positive events lead to the largest relapse probability and
propose interventions to minimize the likelihood of relapse. We find
that the best protective factor is exposure to a mild, yet continuous,
source of contentment, rather than large, episodic jolts of happiness.
\end{abstract}


\end{frontmatter}

\section{Introduction}
Illicit drug abuse remains a major problem in the United
States. Despite decades of research and the implementation of policies
ranging from harm reduction to punitive measures, drug overdose deaths
have increased dramatically over the past 40 years, surpassing 107,000
fatalities in 2022 \cite{Ahmad2022}.  According to the 2021 National
Survey on Drug Use and Health (NSDUH) about 3.3$\%$ of the population
aged 12 and above misused opioids in 2021, the latest year for which
data is available \cite{NSDUH2021}.
	
Our understanding of substance abuse has also evolved in the past 40
years: addiction, once viewed as a lifestyle choice, is now considered
a chronic brain disease characterized by the compulsive seeking and
using of drugs despite harmful consequences.  Drugs change the
neurocircuitry of the brain reward system leading to distortions in
how non-drug rewards are processed, diminished self-control, increased
sensitivity to stressful events, and the prioritization of drug
consumption above all.  Over time, tolerance emerges so that for
pleasurable sensations to persist or for withdrawal symptoms to
dampen, one must increase dosage or intake frequency.  Since
drug-induced damage to the brain is long-lasting and structural,
treatment is a complex process, spanning several years and
necessitating behavioral and pharmacological approaches
\cite{Volkow2015}.  While detoxification requires a few weeks,
remaining sober over a lifetime is challenging: according to the
National Institute of Drug Abuse (NIDA) more than 60$\%$ of those with
substance use disorder relapse within one year \cite{NIDA2020,
  McLellan2000, Sinha2011}.  The likelihood of relapse is highest in
the first months after detoxification \cite{Brecht2014}; however,
relapse is possible even after many years of abstinence
\cite{Smyth2010}.  Since those in recovery may have lost their
previously built tolerance, de-novo consumption, even in smaller
amounts than during active use, may cause overdoses.
 
Given the severity of the problem, it is important to understand the
psychological, behavioral and environmental factors that characterize
drug use \cite{Kaye2023, DOrsogna2023, Caprioli2007}.  Many studies
have been developed over the years to illustrate the process of
addiction, utilizing psychiatric concepts, brain imaging studies, and
behavioral surveys \cite{Volkow2003, Koob2010, Goldstein2011,
  Koob2016, Mollick2020, Chou2022, Changeux2006, Gutkin2012}.
Forecasting tools and data analyses have also been presented
\cite{Jalal2018, Bottcher2023, Bottcher2024}.  There is however no
explicit quantitative framework to describe the cognitive processes
behind relapsing, although the presence of emotional stressors and
sensory cues are known to be major influences \cite{Sinha2001,
  Cahill2016, Koob2020, Nie2021, Koob2022b, Vafaie2022}.
  
Among the most vivid memories of addicts (and former addicts) is the
pleasure associated with the first time drugs were consumed, often the
most euphoric part of the drug-taking experience.  ``Chasing the first
high'' is a common refrain, regardless of how far in the past the
first high occurred. This aligns with the so-called ``peak-end" rule
according to which the memory of a past experience is biased by its
most emotionally intense period (the high in this case), and its
ending \cite{Kahneman2000}.  Other less intense periods, or even the
entire duration of the experience, do not carry as much mnemonic
weight \cite{Fredrickson1993}. Relapses may be triggered by stressful
events that lead to the retrieval of euphoric drug-related memories,
such as the first high, and to the anticipation of future euphoria if
drugs are consumed again \cite{Bornstein2020}.  Drugs are viewed as a
way to alleviate the negative affects induced by current stressors and
to increase short term wellbeing \cite{McCabe2016}.  External cues
such as persons, objects, locations, situations connected to past drug
use may also evoke memories associated with prior drug consumption and
pleasure \cite{Perry2014, Madangopal2019}.  When stressors and/or cues
are present, the associations between drug use and pleasure (or
mitigation of pain) may lead to intense cravings and relapse
\cite{Weiss2001}.  The goal of this work is to create a mathematical
framework whereby the relapse likelihood is described as a function of
quantities that represent life stressors occurring at various times
and with varying intensity, cues and memories related to the previous
drug addiction experience, and changes to the neurocircuitry of the
former user.

In the next section, we introduce our mathematical model in which the
relapse rate is framed in terms of the mental state of the user, drug
availability and the presence of cues. Known psychological and
behavioral processes associated with addiction, such as reward
collection, tolerance, adaptation, and decision-making
\cite{Koob2008,Chou2022,Changeux2006,Gutkin2012,GUTKIN2014,Duka2010}
are integrated into a probabilistic model of relapse events.  Most
critical of these components is a ``mental state'' that is driven by
positive life experiences, stressors and cues. Predictions of our
model, subject to different sequences of positive events, stressors,
and cues are shown in Section \ref{results}. We end with further
discussion and conclusions in Section \ref{conclude}.

\section{Dynamical systems model for relapse}
\label{modelgeneral}

\subsection{Relapse rates and probabilities}

We begin by assuming that drug consumption has ceased and that the
individual started recovery at time $t=0$.  At any time $t>0$ of the
recovery phase, the probability per unit time of relapse, defined as
the instant the individual breaks sobriety by drug intake, is assumed
to be driven by the user's mental state $M(t)$, which can be either
positive or negative, the influence $C(t)$ of any external cues that
remind the user of past drug taking euphoria, and the current
availability of drugs, $I(t)$.  Positive values of the mental state
$M(t)$ indicate well-being and optimism, and negative values represent
discontent and malaise.  Drug availability can be described by a
continuous variable that represents the ease with which drugs are
acquired and consumed.  For simplicity, we binarize $I(t)$ so that
$I(t) =1$ indicates that drugs are readily available and $I(t) =0$
that they cannot be procured.  Finally, cues are assumed to amplify
the relapse rate via a non-negative motivation term $C(t) \geq 0$.
Together, we let $M(t)$, $C(t)$ and $I(t)$ shape the rate of relapse
$R(t)$ via
\begin{equation}
	\label{rate}
	R(t) = I(t) R_0 e^{C(t)} e^{- M(t)}. 
\end{equation}
In this model, $R(t)\dd t$ can be interpreted as the
 probability that the relapse event (first use of drugs after $t=0$)
 occurred between $t$ and $t+ \dd t$. Even though the instantaneous
 relapse rate does not explicitly depend on history or memory, it
 depends on $C(t)$ and $M(t)$ which dynamically evolve, implicitly
 imparting event histories into the current relapse rate.
Eq.\,\ref{rate} indicates that if the drug supply is unrestricted
($I(t)=1$), no cues are present ($C(t) =0$), and an individual is
under a ``neutral" mental state ($M(t)=0$) the rate of relapse is given
by a reference baseline $R(t) = R_0$.  Negative values of the mental
state $M(t) < 0$ increase the relapse rate, conversely $R(t)$ vanishes
in the case of a strongly positive mental state $ M(t) \gg 1$.
An alternative model for $R(t)$ may include a maximal saturated value
$R_{\rm max}$, representing the fastest possible rate of acquiring
 and consuming drugs and that is attained when $C(t)-M(t)$ surpasses a
positive threshold. Finally, the probability of relapsing
by time $T$, $P(T)$, can be written in terms of the survival (against
relapse) probability up to time $T$, $S(T)$ given by
\begin{equation}
	\label{survive}
	S(T) = \displaystyle e^{-\int_0^{T} R (t) {\rm d}t},\quad P(T)= 1 - S(T).
\end{equation}
Next, we describe an event-based model for the dynamics of 
the mental state $M(t)$.
\subsection{The PA/NA mental state model}
\label{mental2}

The so-called ``Positive Activation, Negative Activation'' (PA/NA)
model posits that affects arising from positive and negative
experiences are not coupled \cite{Watson1999, Gross1995, Ito1998b}
and might be processed on different neural substrates
\cite{Cacioppo1999, Lang1995}.  Thus a realistic representation of the
mental state $M(t)$ is as a sum of two contributions, $M(t) = M_{\rm
  a} (t) + M_{\rm b}(t)$, where positive events affect $M_{\rm a}
(t)$, negative ones affect $M_{\rm b}(t)$, and the two evolve
independently.  Negative events, or stressors, are known to impact
one's mental state more than positive ones, a neurological phenomenon
known as the ``negativity bias'' \cite{Ito1998}; recent studies also
show that stressors tend to affect drug users more than the general
population \cite{Zilberman2019, Zilberman2020}, and that drug abuse
produces hypersensitivity to negative emotional distress
\cite{Koob2020, Koob2020b, Koob2021}.  Note that the exponential term
$e^{-M(t)}$ in Eq.\,\ref{rate} weights negative mental states more
than positive ones, in accordance with negativity bias \cite{Ito1998}.
We model the dynamics of $M_{\rm a}$ and $M_{\rm b}$ using different
processing rates $\kappa_{\rm a}(t)$ and $\kappa_{\rm b}(t)$ as

\begin{subequations}
  \begin{align}
	\frac{\dd M_{\rm a}}{\dd t} = & - \kappa_{\rm a}(t) M_{\rm a}+
        \!\sum_{i, t \geq t_{i}^{\rm a}}\! A_{i} \delta(t -t_{i}^{\rm a}), \label{mooda}\\
	\frac{\dd M_{\rm b}}{\dd t} = & - \kappa_{\rm b}(t) M_{\rm b} -
        \!\sum_{j, t \geq t_{j}^{\rm b}}\! B_{j} \delta(t -t_{j}^{\rm b}).\label{moodb}
        \end{align}
\end{subequations}
In Eq.\,\ref{mooda} $A_i > 0 $ is the intensity of positive life event
$i$, as experienced by the individual in recovery, occurring at time
$t_{i}^{\rm a}$ and $\kappa_{\rm a}(t) >0$ is the processing rate that
returns $M_{\rm a}(t)$ to steady state.  Similarly for $- B_j < 0$,
$t_j^{\rm b}$ and $\kappa_{\rm b}(t)$ in Eq.\,\ref{moodb}.  Since
$M_{\rm a}$ and $M_{\rm b}$ are decoupled and $\kappa_{\rm a} (t) \neq
\kappa_{\rm b}(t)$, Eqs.~\ref{mooda} and \ref{moodb} are our
mathematical representation of the PA/NA model.  We solve them
assuming that there are no initial affects, $M_{\rm a} (t=0) = M_{\rm
  b}(t=0)=0$ and that $\kappa_{\rm a}(t) = \kappa_{\rm a}$ and
$\kappa_{\rm b}(t) = \kappa_{\rm b}$ are time-independent.  Non-zero
initial affects can be incorporated by setting $A_1 = M_{\rm a} (t=0)$
at $t_{i=1}^{\rm a} = 0$ or $B_1 = - M_{\rm b}(t=0)$ at $t_{j=1}^{\rm
  b} = 0$ in the sequence of positive or negative life events.
Time-dependent $\kappa_{\rm a}(t), \kappa_{\rm b}(t)$ are discussed in
the Appendix.  We solve Eqs.~\ref{mooda} and \ref{moodb} under the
above approximations to find

\begin{figure*}[t!]
\centering
  \includegraphics[width=0.75\linewidth]{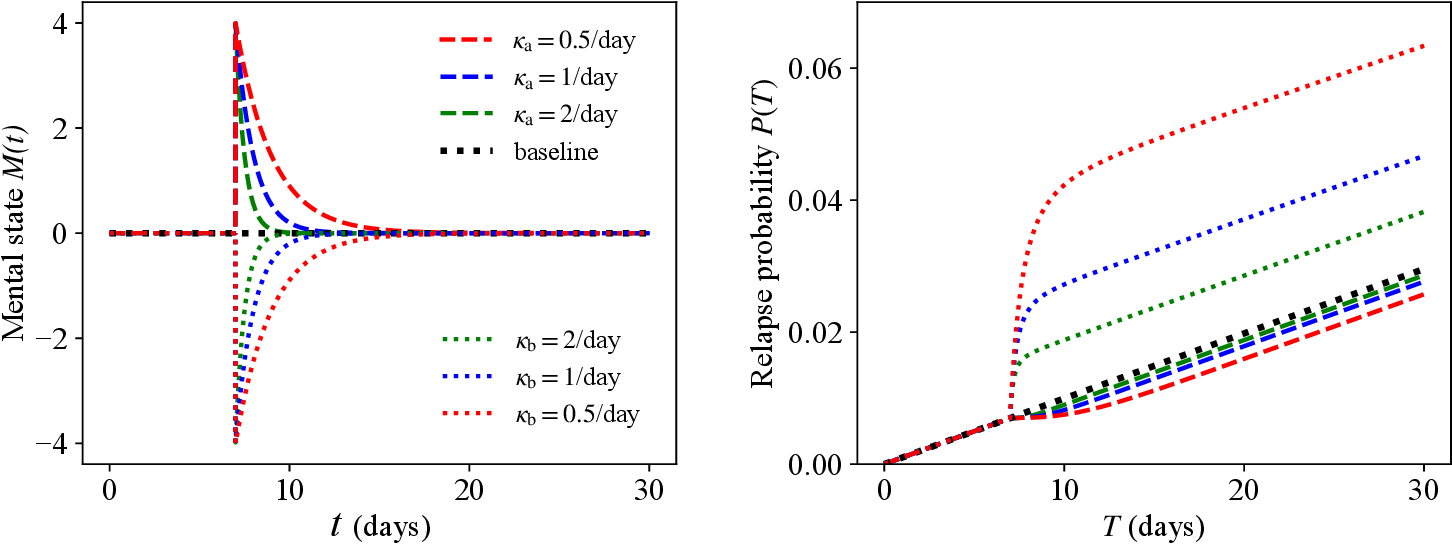}
\caption{Mental state at time $t$, $M(t)$, and the probability of
  relapse before time $T$, $P(T)$, upon exposure to a single stressor
  $\{- B_1, t_1^{\rm b}\}$ or to a single positive event $\{A_1,
  t_1^{\rm a}\}$ for three processing rates $\kappa_{\rm b} \text{ or
  } \kappa_{\rm a} =2, 1, 0.5$/day with $B_1=A_1=4$, $t^{\rm b}_1,
  t^{\rm a}_1=7$ days, $R_0=10^{-3}$/day ,$~M_0 = 0$.  The relapse
  probability decreases with $\kappa_{\rm b}$ so that the longer a
  stressor impacts one's mental state, the larger the likelihood of
  relapse. The opposite is true for positive events, for which the
  longer memory of a positive event is retained, the less likely
  relapse is. Note the more pronounced effect of the negative mental
  state $B_1$, compared to the positive one $A_1$ under the same
  processing rate despite their amplitudes being the same.  The mental
  state $M(t) = M_{\rm a}(t)+ M_{\rm b}(t)$ is given by
  Eqs.\,\ref{Ma_Mb}; the relapse probability $P(T)$ by 
  Eqs.~\ref{rate} and \ref{survive}.}
\label{fig:unchange kappa}
\end{figure*}

\begin{equation}
\begin{aligned}
	M_{\rm a}(t) = & \!\sum_{i, t \geq t_i^{\rm a}}\! A_i e^{-\kappa_{\rm a} (t-t_i^{\rm a})}, \\
	M_{\rm b}(t) = & -\!\sum_{j, t \geq t_j^{\rm b}}\! B_j e^{-\kappa_{\rm b} (t-t_j^{\rm b})}. 
\label{Ma_Mb}
\end{aligned}
\end{equation}
The mental state integrated up to time $T$ after $n_{\rm a}$ positive and
$n_{\rm b}$ negative life events, such that $t_{n_{\rm a}}^{\rm a} \leq T
\leq t_{n_{\rm a}+1}^{\rm a}$ and $t_{n_{\rm b}}^{\rm b} \leq T \leq
t_{n_{\rm b}+1}^{\rm b}$, is thus given by
\begin{equation}
\begin{aligned}
\label{overall}
\int_0^T \!M(t)\dd t = & \int_0^T \!\big(M_{\rm a}(t)
+ M_{\rm b}(t)\big) \dd t \\
	\: = & \frac{1}{\kappa_{\rm a}} 
        \sum_{i=1}^{n_{\rm a}}\! A_i \left(1- e^{-\kappa_{\rm a}(T- t_i^{\rm a})}\right)
        - \frac{1}{\kappa_{\rm b}}\sum_{j=0}^{n_{\rm b}} \! B_j
        \left(1- e^{- \kappa_{\rm b}(T- t_j^{\rm b})}\right). 
\end{aligned}
\end{equation}
The effects of a sequence of $n_{\rm a}$ events defined by $\{A_i,
t_i^{\rm a} \}$ on the integrated mental state in Eq.\,\ref{overall}
can be reproduced by a single event of amplitude $Z_{\rm a}$ at specific time
$t_{\rm a}$
\begin{equation}
	\label{equiv_A}
	Z_{\rm a}=\sum_{i=1}^{n_{\rm a}} A_i, \,\,\,
	t_{\rm a} = \frac{1}{\kappa_{\rm a}}\ln
        \left[\displaystyle{\frac{\sum_{i=1}^{n_{\rm a}} A_i e^{\kappa_{\rm a} t_i^{\rm a}}}
            {\sum_{i=1}^{n_{\rm a}} A_i}}\right].
\end{equation}
\noindent
Similarly, a sequence of $n_{\rm b}$ 
events $\{ - B_j, t_j^{\rm b} \}$ generates an integrated mood
that can be reproduced by single event 
\begin{equation}
	\label{equiv_B}
	Z_{\rm b} = \sum_{j=1}^{n_{\rm b}} B_j, \,\,\,
	t_{\rm b} = \frac{1}{\kappa_{\rm b}}\ln
	\left[ \displaystyle{\frac{\sum_{j=0}^{n_{\rm b}}
              B_j e^{\kappa_{\rm b} t_j^{\rm b}}}{\sum_{i=0}^{n_{\rm b}} B_j}}\right].
\end{equation}
Thus far, we have modeled the dynamics of positive and negative mental
state variables. Included in the relapse rate $R(t)$ is also a
dependence on random cues that trigger the memory of drug-induced
euphoria. The model for cues shares many features of the negative mood
variable and is described below.

\subsection{External cues}
Here we discuss representations for $C(t) \geq 0$. External cues can
trigger memories of the pleasurable feelings associated with drug
taking \cite{Sinha2007, Fox2007}. We model these memories as impulses
occurring at times $t_{\ell}^{\rm c}$ whose effects decay with rate
$\kappa_{\rm c}$.  Thus, the dynamics of $C(t)$, the overall motivation
from cues, is given by
\begin{equation}
	\label{cue}
	\frac{\dd C}{\dd t} =
        -\kappa_{\rm c}(t) C + \!\sum_{\ell, t \geq t_\ell^{\rm c}}\!
        C_{\ell} \delta(t -t_{\ell}^{\rm c}).
\end{equation}
The amplitude $C_{\ell}$ represents the mnemonic strength of a given
cue.  By the peak-end rule, we assume that the most intense memory is
proportional to $w_{\rm peak} > 0$, the largest reward response during
addiction, and set $C_{\ell} = w_{\rm peak}$ for all $\ell$.  We also
assume the decay rate $\kappa_{\rm c}(t)$ associated with the
permanence of the cue in one's memory is a constant, $\kappa_{\rm
  c}(t) = \kappa_{\rm c}$ and that there are no initial cues, $C(t=0)
= 0$, which leads to
\begin{equation}
\label{wpeak}
	C(t) = w_{\rm peak} \sum_{\ell, t \geq t_{\ell}^{\rm c}}  e^{- \kappa_{\rm c} (t- t^{\rm c}_{\ell})}.
\end{equation}

\section{Results} 
\label{results}

\begin{figure*}[t]
  \centering
\includegraphics[width=0.75\linewidth]{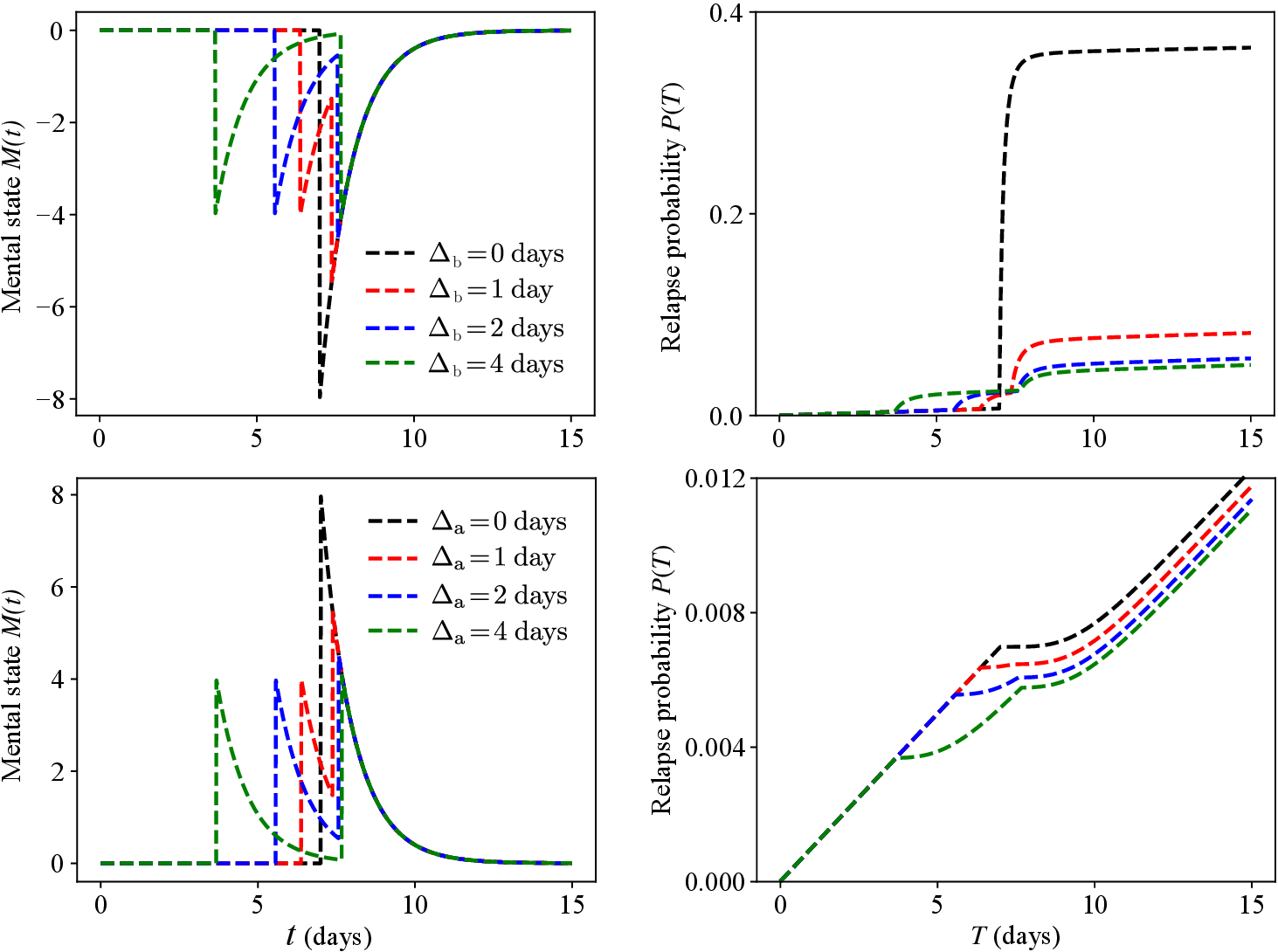}
\caption{Top row: Mental state $M(t)$ and relapse probability $P(T)$
  upon exposure to two stressors $\{-B_1, t_1^{\rm b} \}$ and $\{-B_2,
  t_2^{\rm b}\}$ separated by lag times $\Delta_{\rm b} = t_2^{\rm b}
  - t_1^{\rm b} = 0, 1, 2, 4$ days and obtained using
  Eqs.\,\ref{rate}, \ref{survive}, \ref{Ma_Mb}, and \ref{cluster2}.
  Parameters are $\kappa_{\rm b}= 1$/day, $B_1=B_2=4$,
  $~R_0=10^{-3}$/day,$~M_0 = 0$.  All stressor pairs define the same
  integrated mental state defined in Eq.\,\ref{overall} and are
  equivalent to the single event $\{-Z_{\rm b}, t_{\rm b} \}$ shown in
  the red curve.  For each $\Delta_{\rm b}$, the corresponding
  $t_1^{\rm b}$ is derived from the constraint $H_{\rm b} =
  B_1e^{\kappa_{\rm b }t_1^{\rm b}}+B_2e^{\kappa_{\rm b} (t_1^{\rm b}
    + \Delta_{\rm b})}$ where $H_{\rm b} = Z_{\rm b} e^{ \kappa_{\rm
      b} t_{\rm b}}$, $Z_{\rm b} = (B_1 + B_2)$, and $t_{\rm b} = 7$
  days.  Notice that $P(T)$ decreases with $\Delta_{\rm b}$, implying
  that stressors should be as spaced apart as possible to decrease the
  likelihood of relapse in accordance with our analytical findings.
  Bottom row: Corresponding plots for two positive events, $\{A_1,
  t_1^{\rm a} \}$ and $\{A_2, t_2^{\rm a}\}$ separated by lag times
  $\Delta_{\rm a} = t_2^{\rm a} - t_1^{\rm a} = 0, 1, 2, 4$ days with
  $A_1 = A_2 = 4$ and all other parameters the same as above.  The
  constraint can be obtained by setting ${\rm a} \to {\rm b}$ in the
  two-stressor constraint expression, with $t_{\rm a} = 7$ days.
  Here, $P(T)$ decreases with $\Delta_{\rm a}$, implying that the best
  protection against relapse is by experiencing well-spaced positive
  events rather than large clustered ones. Small, repeated joys and
  small, repeated unpleasant events are better than large a jolt of
  happiness or catastrophe.}
\label{fig:2events}
\end{figure*}
 
 \begin{figure*}[t!]
\centering
   \includegraphics[width=0.75\linewidth]{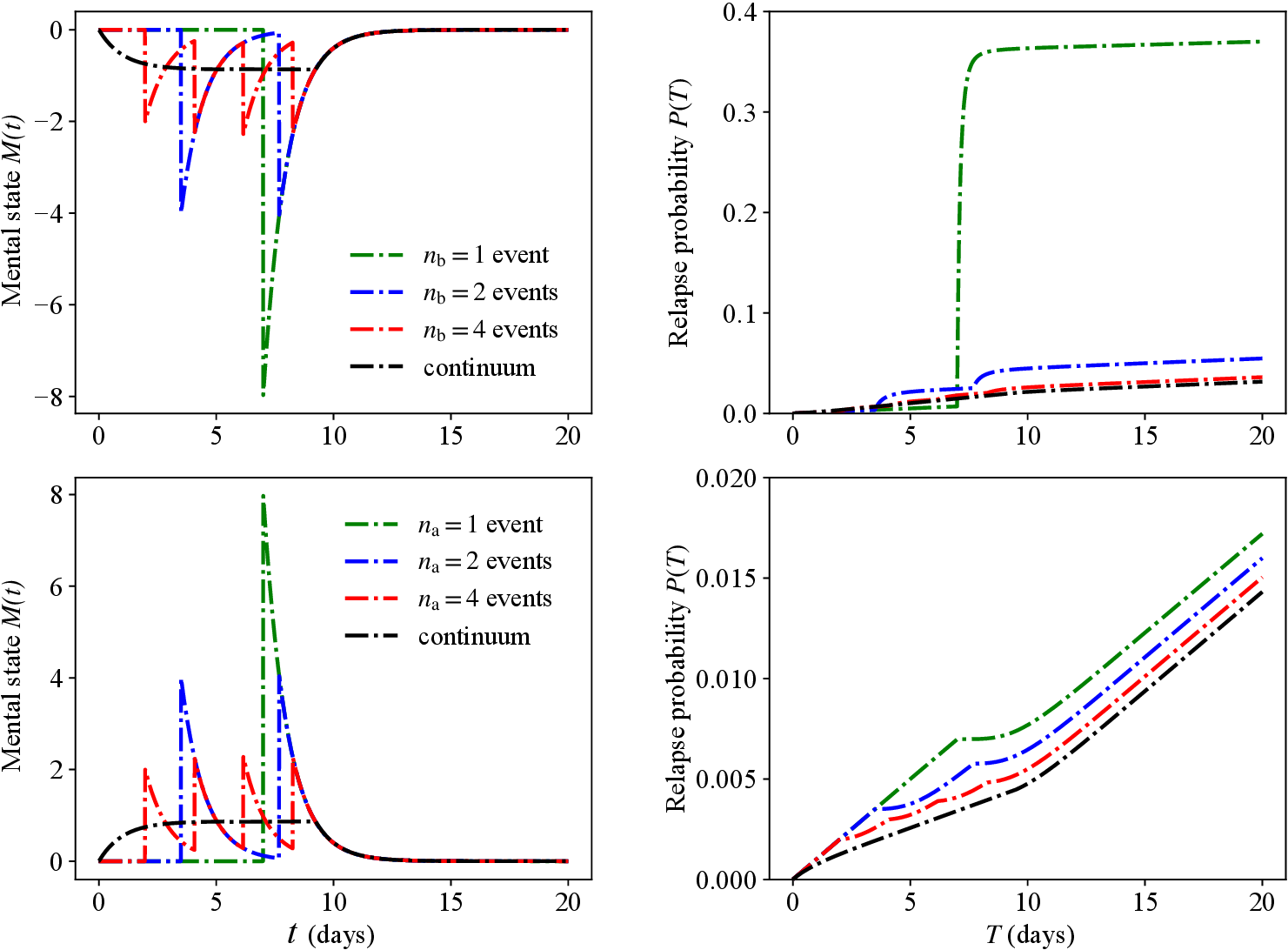}
\caption{Top row: Mental state $M(t)$ and relapse probability $P(T)$
  upon exposure to $n_{\rm b}=1,2,4$ negative events $\{-B_i,
  t_{i}^{\rm b} \}$ and a continuum of stressors. All curves are
  obtained using Eqs.\,\ref{rate}, \ref{survive}, \ref{Ma_Mb}, and
  \ref{cluster2}.  Sequences define the same integrated mental state
  defined in Eq.\,\ref{overall} and are equivalent to the single event
  $\{-Z_{\rm b}, t_{\rm b} \}$ shown in the red curve, where $Z_{\rm
    b} = 8$ and $t_{\rm b} = 7$ days.  Within each sequence, events
  carry the same amplitude $B_i$ and are separated by the same time
  interval $\Delta_{\rm b}=$ 3.5 days ($n_{\rm b}=2$, black curve),
  2 days ($n_{\rm b}=4$, blue curve), and 0.1 days (continuum, green
  curve).  Other parameters are $\kappa_{\rm
    b}=1$/day,$~R_0=10^{-3}$/day,$~M_0 = 0$.  Bottom row:
  Corresponding plots for sequences of equivalent positive events with
  $Z_{\rm a} = 8$ and $t_{\rm a} = 7$ days with $\kappa_{\rm a} =
  1$/day and $\Delta_{\rm a} =$ 3.5 days ($n_{\rm a}=2$, black
  curve), 2  days ($n_{\rm a}=4$, blue curve), and 0.1 days (continuum,
  green curve).  Notice that for both positive and negative events the
  relapse probability is lowest for small events of limited
  magnitude. }
\label{fig:small events}
\end{figure*}

We now study how external stimuli and intrinsic traits affect the
relapse probability $P(T)$.  External factors include specific
realizations of the $\{ A_i, t_i^{\rm a} \}$ and $\{ -B_j, t_j^{\rm b}
\}$ sequences, cue occurrence times $\{t_{\ell}^{\rm c}\}$ and drug
availability profile $I(t)$.  The intrinsic characteristics of an
individual include how his or her mental state is affected by
stressors, joyous events, and cues, the processing rates for positive
and negative events, $\kappa_{\rm a}$ and $\kappa_{\rm b}$, for cues,
$\kappa_{\rm c}$, and the intensity of the first high $w_{\rm peak}$.
Relevant parameter ranges are listed in Table \,\ref{tablepar}.
Specifically, we measure time in units of days and fix $R_0 =
10^{-3}$/day consistent with known relapse rates of roughly 40 to 60
percent among opioid abuse disorder patients one year after treatment
\cite{Kampman2015}.  We assume drugs are always available and set
$I(t) =1$ throughout the remainder of this work. Relapse is not
possible if drugs are not available.

\subsection{Dynamics without cues}
\label{nocues}

We begin by studying the case of no cues and an unimpeded drug supply
so that $C(t) =0$ and $I(t) =1$.  We first analyze the simple case of
a single life event
and later consider sequences of multiple negative and positive ones.  Our
goal is to identify which combination of events (intensity and
timing) leads to the smallest 
relapse likelihood.

\subsubsection{Longer-lasting stressors increase the relapse probability; 
longer-lasting positive events decrease it}
\label{nocues1}

We consider a single stressor that is processed at three different
rates $\kappa_{\rm b}$.  Fig.\,\ref{fig:unchange kappa} shows that the
longer the stressor affects one's mental state (\textit{i.e.} the
lower $\kappa_{\rm b}$), the larger the relapse probability $P(T)$.
Corresponding results are shown for a positive experience processed at
various rates $\kappa_{\rm a}$: lower values of $\kappa_{\rm a}$
result in smaller relapse likelihoods, as the effects of the positive
event are retained for a longer time.  Due to the exponential term in
the relapse rate, stressors result in higher relapse likelihoods
compared to positive ones of the same amplitude.
\subsubsection{Clustered stressors increase the relapse probability more
    than disperse ones}
\label{nocues2}

\begin{figure*}[t!]
\centering
  \includegraphics[width=0.75\linewidth]{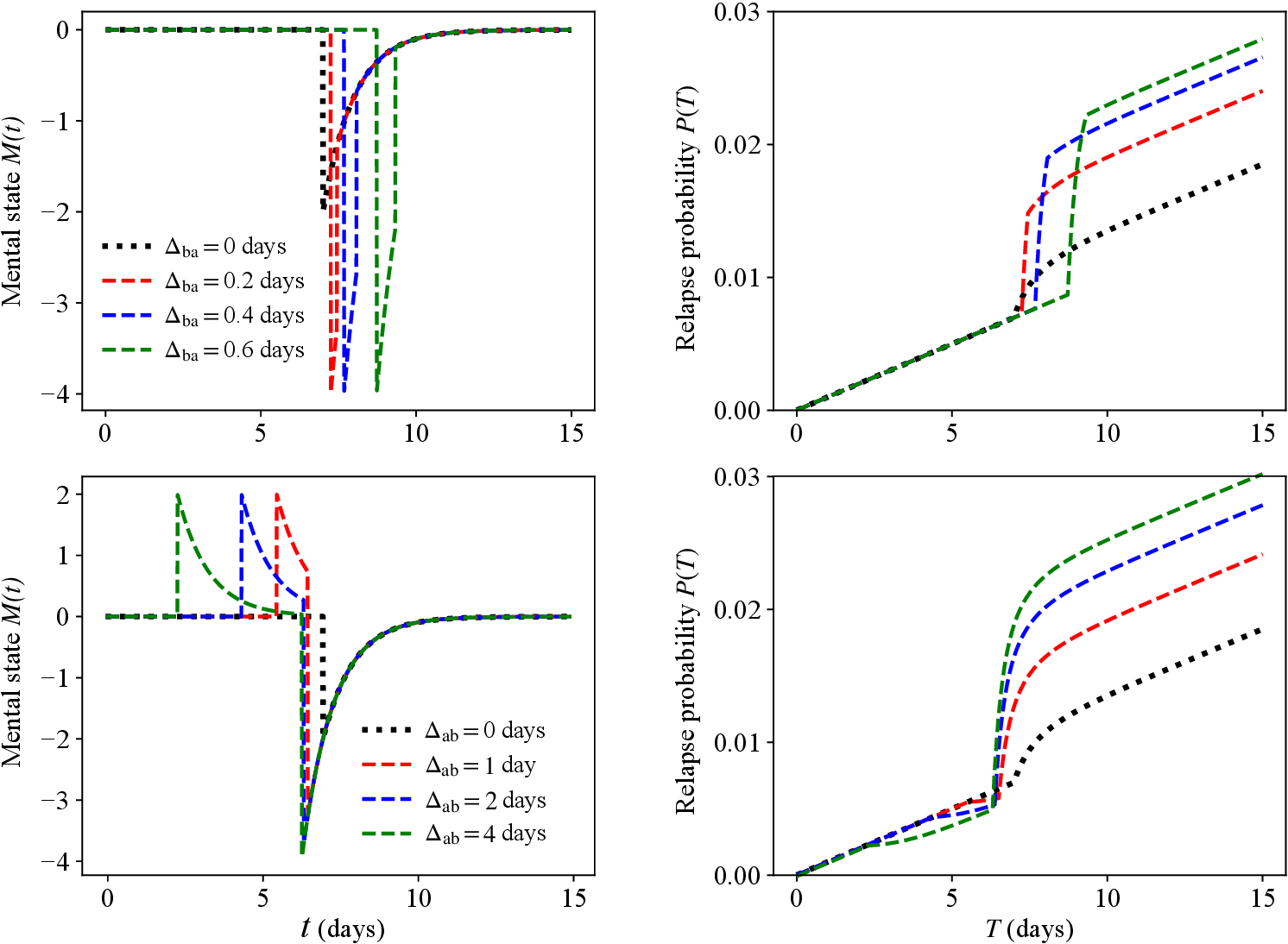}
\caption{Top row: Mental state $M(t)$ and relapse probability $P(T)$
  upon exposure to a stressor $\{-B_1, t_1^{\rm b} \}$ followed by a
  positive event $\{A_2, t_2^{\rm a}\}$. The two events are separated
  by lag times $\Delta_{\rm ba} = t_2^{\rm a} - t_1^{\rm b} = 0, 0.2,
  0.4, 0.6$ hours.  All curves are obtained using Eqs.\,\ref{rate},
  \ref{survive}, \ref{Ma_Mb}, and \ref{dual}.  Parameters are
  $\kappa_{\rm b}=\kappa_{\rm a}= \kappa =1$/day, $B_1=4$, $A_2 = 2$,
  $~R_0=10^{-3}$/day,$~M_0 = 0$ and by setting $H = 2 e^7$.  All event
  pairs define the same integrated mental state defined in
  Eq.\,\ref{overall} for $T > t_2^{\rm a}$.  Notice that $P(T)$
  increases with $\Delta_{\rm ba}$, implying that given a stressor,
  the likelihood of relapse is lowest the earlier a counteracting
  positive event occurs, in accordance with our analytical findings.
  Bottom row: Corresponding plots for a positive event $\{A_1,
  t_1^{\rm a} \}$ followed by a stressor $\{-B_2, t_2^{\rm b}\}$ with
  lag times $\Delta_{\rm ab} = t_2^{\rm b} - t_1^{\rm a} = 0, 1, 2, 4$
  days with $A_1 = 2$, $B_2 = 4$ and all other parameters are the same
  as above.  Here, $P(T)$ also increases with $\Delta_{\rm ab}$ and is
  smallest for $\Delta_{\rm ab} \to 0$.}
\label{fig:ABBA_samek}
\end{figure*}	

We now include multiple life events and study how their timing affects
the likelihood of relapse. Let us start with two negative events,
$\{-B_1, t_1^\mathrm{b}\}$ and $\{-B_2, t_2^\mathrm{b}\}$, that define
the time interval $\Delta_{\rm b}=t_2^{\mathrm{b}}-t_1^{\mathrm{b}}
\geq 0$.  We then define the effective time $U(T)$ given by
\begin{equation}
  \begin{aligned}
\label{cluster}
U(T) & \equiv - \frac{\ln S(T)}{R_0}  =  \int_0^T  e^{-M(s)} \dd s \\
\: & = t_1^{\rm b} + \! \int_{0}^{\Delta_{\rm b}}  e^{B_1 e^{-\kappa_{\rm b}s}} \dd s
+ \! \int_{\Delta_{\rm b}}^{T-t_1^{\rm b}} \!\!
e^{\left(B_1 + B_2 e^{\kappa_{\rm b} \Delta_{\rm b}}\right) e^ {-\kappa_{\rm b}s}} \dd s.
\end{aligned}
\end{equation}
If $B_1, B_2 = 0$ (as in the ``neutral'' case or baseline)
Eq.~\ref{cluster} gives $U(T) = T$; finite values of $B_1, B_2$ lead
to $U(T) > T$, increasing the relapse probability $P(T) = 1 - S(T)$
above that of the baseline.

We now consider the family of paired events where the amplitudes $B_1,
B_2$ are fixed and where $t_1^{\rm b}, t_2^{\rm b}$ are chosen such
that for $T > t_2^{\rm b}$ the two events yield the same integrated
mood as the single event $\{Z_{\rm b}, t_{\rm b} \}$ defined in
Eqs.~\ref{equiv_B}.  This implies that $B_1 e^{\kappa_{\rm b} t_1^{\rm
    b}} + B_2 e^{\kappa_{\rm b} t_2^{\rm b}} \equiv H_{\rm b}$ must be
a constant, leaving one degree of freedom, which we choose to be
$\Delta_{\rm b}$.  The above constraints also impose that the
integrated mental state, $\int_0^{T} M(s) \dd s$ is invariant for all
paired events within the family defined by $B_1, B_2, H_{\rm b}$. We
may now ask: within this family of paired events, where the integrated
mental state is fixed, which choice of $\Delta_{\rm b}$ minimizes the
relapse probability at any time $T > t_2^{\rm b}$?  We first express
$t_1^{\rm b}$ in terms of $H_{\rm b}$ and $\Delta_{\rm b}$,
\begin{equation}
  t^{\rm b}_1 = \frac {1}{\kappa_{\rm b}} \ln
  \bigg(\frac{H_{\rm b}}{B_1 + B_2 e^{\kappa_{\rm b} \Delta_{\rm b}}}\bigg), 
\end{equation}
and make the dependence of $U(T)$ on $\Delta_{\rm b}$ explicit so that
$U(T) \to U(T; \Delta_{\rm b})$ and
\begin{equation}
  \begin{aligned}
    \label{cluster2}
    \kappa_{\rm b}U(T; \Delta_{\rm b}) = & \ln
    \Big(\frac{H_{\rm b}}{B_1 + B_2 e^{\kappa_{\rm b} \Delta_{\rm b}}}\Big) \\
    \: & \,\, + \mbox{ Ei}(B_1) -\mbox{Ei}\big(B_1 e^{-\kappa_{\rm b} \Delta_{\rm b}}\big) \\
    \: & \,\, + \mbox{ Ei}\big(B_1 e^{-\kappa_{\rm b} \Delta_{\rm b}} +B_2\big)
      - \mbox{Ei}\big(H_{\rm b} e^{-\kappa_{\rm b} T}\big)
  \end{aligned}
  \end{equation}
where the exponential integral $\mbox{Ei}(x)\equiv \int_{-\infty}^x
t^{-1} e^t \dd t$. Since $B_1, B_2, H_{\rm b}$ are fixed, the extrema of
$U(T; \Delta_{\rm b})$ with respect to $\Delta_{\rm b}$ at any time
$T$ are given by the zeros of

\begin{equation}
\label{derivative}
\kappa_{\rm b}\frac{ \partial U(T; \Delta_{\rm b})}{\partial \Delta_{\rm b}} =
(1-B_{2})e^{B_1 e^{- \kappa_{\rm b} \Delta_b}}\! +
\frac{e^{B_{2}}e^{B_1 e^{- \kappa_{\rm b} \Delta_b}}\!-1}{1+\frac{B_{1}}{B_{2}}
  e^{-\kappa_{\rm b} \Delta_{\rm b}}}.
\end{equation}
Regardless of $B_1, B_2, \kappa_{\rm b}$, the left hand side of
Eq.~\ref{derivative} is a negative function of $\Delta_{\rm b}$,
implying that $U(T;\Delta_{\rm b})$ has a maximum at $\Delta_{\rm b}
\to 0$.  Since $P(T) = 1 - S(T) = 1 - e^{-R_0 U(T;\Delta_{\rm b})}$,
we conclude that the largest relapse probability also occurs at
$\Delta_{\rm b} \to 0$; that is, the relapse likelihood is largest
when the two negative events occur simultaneously.  In the top row of
Fig.\,\ref{fig:2events} we show the relapse probability $P(T)$ upon
exposure to pairs of stressors that belong to the same family of
events with fixed $B_1, B_2, H_{\rm b}$ and different timings
$\Delta_{\rm b}$.  The relapse probability is indeed largest when the time lag
between the two stressors is smallest, $\Delta_{\rm b} \to 0$.  We can
apply the same arguments to more than two events and show that as the
number of stressors increases, so does the likelihood of relapse.
Given a negative integrated mental state generated by a set of $n_{\rm
  b}$ negative events occurring within time $T$, the relapse
likelihood is largest when stressors are coincident, and is reduced
when stressors are spread out.  Fig.\,\ref{fig:small events} shows the
relapse probability for $n_{\rm b} = 1,2,4$ and a continuum of
negative events.  In our model, relapse after a single catastrophic
event is more likely than after a series of smaller stressors which
cumulatively yield the same integrated negative mental state as the
single catastrophic stressor.
\begin{figure*}[htb]
  \centering
	\includegraphics[width=0.75\linewidth]{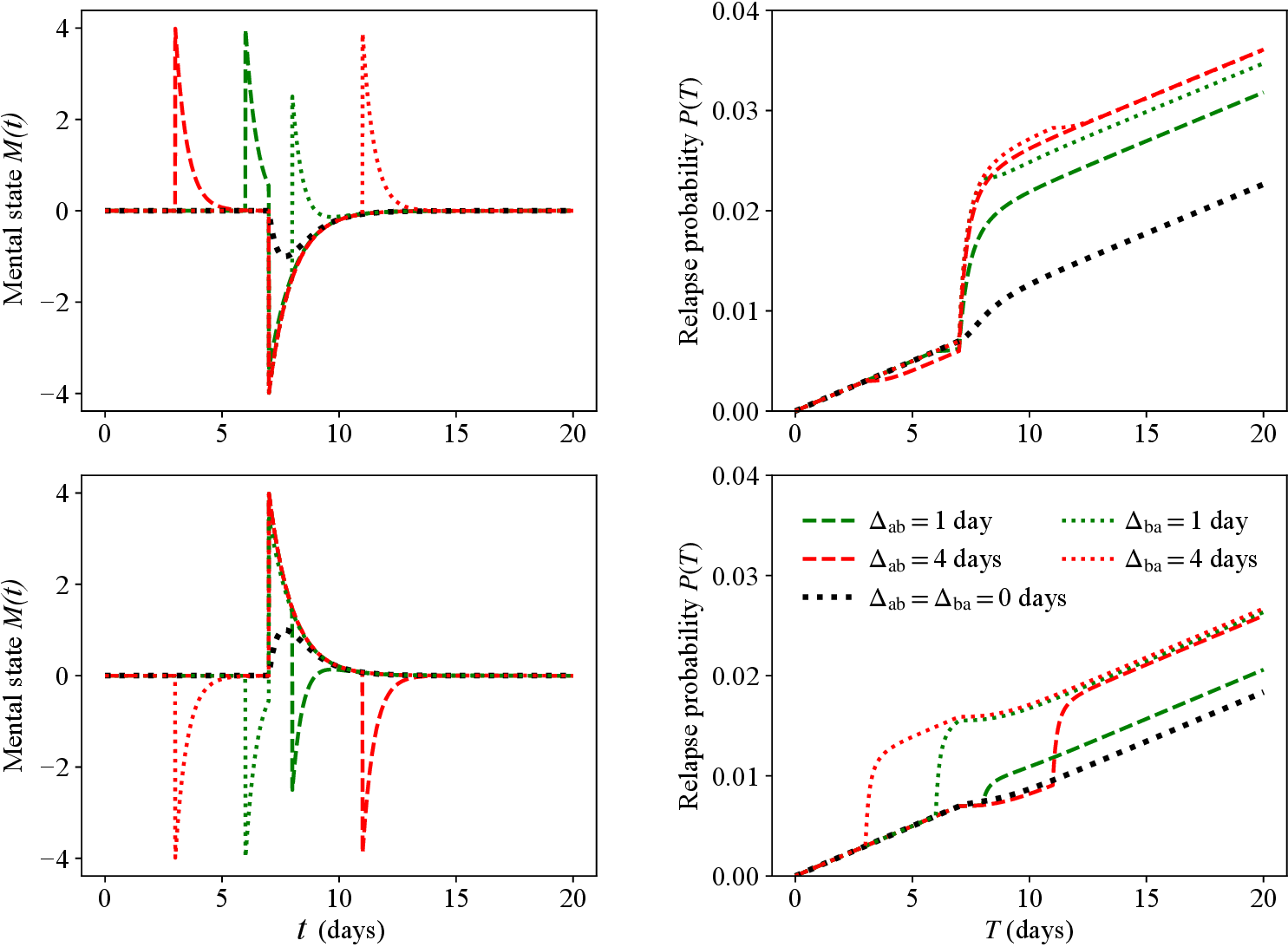}
	\caption{Top row: Mental state $M(t)$ and relapse probability
          $P(T)$ upon exposure to two events of
          opposite sign $\{A_i, t_i^{\rm a} \}$ and $\{-B_j, t_j^{\rm
            b} \}$ for $\kappa_{\rm a}=2$/day and $\kappa_{\rm
            b}=1$/day where $i=1, j=2$ or $j=1, i=2$. All curves are
          obtained using Eqs.\,\ref{rate}, \ref{survive}, \ref{Ma_Mb},
          and \ref{dual}.  Events carry the amplitudes $A_i= B_j=4$;
          the other parameters are set at $M_0=0,R_0=10^{-3}$.  Lag
          times are evaluated by setting $t_j^{\rm b} = 7$ days where
          $j=1$ for lag times $\Delta_{\rm ba} = t_2^{\rm a} -
          t_1^{\rm b}$ and $j=2$ for lag times $\Delta_{\rm ab} =
          t_2^{\rm b} - t_1^{\rm a}$.  For large enough $\Delta_{\rm
            ab} = \Delta_{\rm ba} = 4$ days the relapse probability is
          independent of the order of events; however, for smaller lag
          times $\Delta_{\rm ab} = \Delta_{\rm ba} = 1$ day, the order
          of events matters and the relapse probability is lower when
          the positive event $\{A_1, t_1^{\rm a} \}$ occurs prior to
          the stressor $\{-B_2, t_2^{\rm a} \}$.  If the two events
          are concurrent and $\Delta_{\rm ab} = \Delta_{\rm ba} = 0$
          the total input to the mental state is negative since
          $\kappa_{\rm b} < \kappa_{\rm a}$ and the effects of the
          stressor are retained for a longer time than those of the
          positive event.  Bottom row: Corresponding plots for
          $\kappa_{\rm a}=1$/day and $\kappa_{\rm b}=2$/day with all
          other parameters and events the same as those of the top
          row, and where the ordering of positive and negative events
          is reversed.  Here, the order of events plays an even more
          crucial role when the time lag $\Delta_{\rm ab} =
          \Delta_{\rm ba} = 1$ as the relapse probability is much
          lower when the positive event $\{A_1, t_1^{\rm a} \}$ occurs
          prior to the stressor $\{-B_2, t_2^{\rm a} \}$.  Since
          $\kappa_{\rm b} \neq \kappa_{\rm a}$ it is not possible for
          the pairs of events to define the same mental state for all
          values of $t$.}
	\label{fig:ABBA_ka_kb}
\end{figure*}

\subsubsection{Dispersed positive events decrease the relapse
  probability more than clustered ones}
\label{nocues3}
We can derive similar expressions to Eq.\,\ref{cluster} for two
positive life events for which the corresponding $U(T)$ is obtained by
substituting $B_{1,2} \to -A_{1,2}$, and $t_1^{\rm b}, \Delta_{\rm b}, 
\kappa_{\rm b} \to t_1^{\rm a}, \Delta_{\rm a}, \kappa_{\rm a}$,
respectively. Using the same methods as for pairs of negative events,
we can show that the occurrence of two positive events decreases the
likelihood of relapse the most when the two events are well spaced
out.  Results are shown in the bottom row of Fig.\,\ref{fig:2events}
for pairs of positive events with fixed $A_1, A_2, H_{\rm a}$ and
different timings. The smallest likelihood of relapse occurs when the
time lag between positive events $\Delta_{\rm a}$ is large,
conversely, the likelihood of relapse is most pronounced for
$\Delta_{\rm a} \to 0$.  In the case of multiple, positive life
events, as with the findings described for negative life events, the
likelihood of relapse decreases the most when an individual
experiences many distributed but moderately happy events, compared to
a much larger but isolated positive episode that carries the same
overall impact as the distributed ones.  Results for sequences of
multiple events belonging to the same family are shown in the bottom
row of Fig.~\ref{fig:small events} for $n_{\rm a} = 1,2,4$ where
events are equally spaced and for a continuum of episodes.  As
expected, the lowest relapse probability occurs for a uniform
distribution of positive events.  A modest but continuous source of
support is more protective against relapse than a very intense yet
short-lived positive experience.

\subsubsection{Relapse is least likely if 
a positive experience occurs immediately after a stressor}
\label{nocues4}

We now examine the case of a stressor $\{- B_1, t_1^{\rm b} \}$
followed by a positive event $ \{A_2, t_2^{\rm a} \}$, where $t_2^{\rm
  a} > t_1^{\rm b}$. We label the positive event $A_2$ rather than
$A_1$ so that it is clear that the positive event occurs after the
negative one. Given an integrated mental state $\int_{0}^T M(t') \dd
t'$ and values for $B_1, A_2$, the goal is to establish the lag time
between the two events that minimizes the likelihood of relapse.  The
general case of different processing rates $\kappa_{\rm a} \neq
\kappa_{\rm b}$ does not allow for easy generalization, so we set
$\kappa_{\rm a} = \kappa_{\rm b} = \kappa$ to simplify our
analysis. We write $U(T) = - \ln S(T)/R_0 $ in terms of $\Delta_{\rm
  ba} = t_2^{\rm a} - t_1^{\rm b}$ and make the dependence on
$\Delta_{\rm ba}$ explicit in the expression for $U(T; \Delta_{\rm
  ba})$ as follows
\begin{equation}
  \begin{aligned}
  \kappa U_{\rm}(T; \Delta_{\rm ba} ) = & \ln
  \bigg(\frac{H}{B_1 - A_2 e^{\kappa \,\Delta_{\rm ba}}}\bigg)\\
  \: & \,\, + \mbox{ Ei} (B_1) - \mbox{Ei} (B_1 e^{-\kappa \Delta_{\rm ba}}) \\
\: & \,\, + \mbox{ Ei} (B_1 e^{-\kappa \Delta_{\rm ba}} -A_2) - \mbox{Ei}(H e^{-\kappa  T}),
  \end{aligned}
  \label{dual}
\end{equation}
where $H \equiv B_1 e^{\kappa t_1^{\rm b}} - A_2 e^{\kappa t_1^{\rm
    a}}$ is a constant that ensures $\int_{0}^T M(t') \dd t'$ is
independent of $\Delta_{\rm ba}$.  We can thus write

\begin{equation}
\begin{aligned}
\kappa  \frac{ \partial U (T; \Delta_{\rm ba})}{\partial \Delta_{\rm ba}}  = &
\, A_2 e^{B_1 e^{- \kappa \Delta_{\rm ba}}} \\
\: & \,\, \times \Big(
\frac{1 - e^{-\kappa A_2}}{A_2} +
\frac {e^{-B_1 e^{-\kappa \Delta_{\rm ba}}} - e^{-\kappa A_2}}{B_1 e^{-\kappa \Delta_{\rm ba}} - A_2}\Big).
\end{aligned}
\label{derivative2}
\end{equation}
The left hand side of Eq.\,\ref{derivative2} is a positive function of
$\Delta_{\rm ba}$ regardless of $A_1, B_2, \kappa$; as a consequence,
for $T > t_2^{\rm a}$, $U(T; \Delta_{\rm ba})$ is an increasing
function of $\Delta_{\rm ba}$ and attains its lowest value at
$\Delta_{\rm ba} = 0$. Thus, the relapse probability is also the
lowest for $\Delta_{\rm ba} = 0$. Once a negative life event occurs,
the way to minimize the occurrence of relapses is for the individual
to experience a healing, positive experience as soon as possible.
Similar results hold in the case of a positive event $\{A_1, t_1^{\rm
  a} \}$ followed by a negative event $\{B_2, t_2^{\rm b} \}$; the
relapse probability is lowest when the time lag between the two events
is shortest.

In Fig.\,\ref{fig:ABBA_samek} we show the relapse probability for
equivalent pairs of events $\{-B_1, t_1^{\rm b} \}$ and $\{A_2,
t_2^{\rm a} \}$ that define a fixed $H \equiv B_1 e^{\kappa t_1^{\rm
    b}} - A_2 e^{\kappa t_2^{\rm a}}$ for $\kappa_{\rm a} =
\kappa_{\rm b} = \kappa$. For large enough $T$, $P(T)$ decreases with
$\Delta_{\rm ba}$, confirming our analytic predictions. The same
finding arises for equivalent pairs of events $\{A_1, t_1^{\rm a} \}$
and $\{-B_2, t_2^{\rm b} \}$ that define a fixed $H \equiv
-A_1e^{\kappa t_1^{\rm a}} + B_2 e^{\kappa t_2^{\rm b}}$ for
$\kappa_{\rm a} = \kappa_{\rm b} = \kappa$.  To derive the
mathematical results presented for pairs of events of amplitude $(A_1,
A_2)$, $(B_1, B_2)$, $(B_1, A_2 )$ and $(A_1, B_2)$ we assumed that no
other prior events occurred; however, it is possible to show that our
findings remain valid in the presence of earlier events.  For example,
given the events $\{A_1, t_1^{\rm a} \}$, $\{-B_2, t_2^{\rm b} \}$,
$\{-B_3, t_3^{\rm b} \}$ with $t_3^{\rm b} \geq t_2^{\rm b} \geq
t_1^{\rm a}$ processed at the same rate $\kappa_{\rm a} = \kappa_{\rm
  b} = \kappa$ one can show that the relapse probability is still
maximized upon clustering the two negative events $(B_2, B_3)$ and
setting $\Delta_{\rm b} = t_3^{\rm b} - t_2^{\rm b} \to 0$.

\begin{figure*}[htb]
\centering
\includegraphics[width=0.86\linewidth]{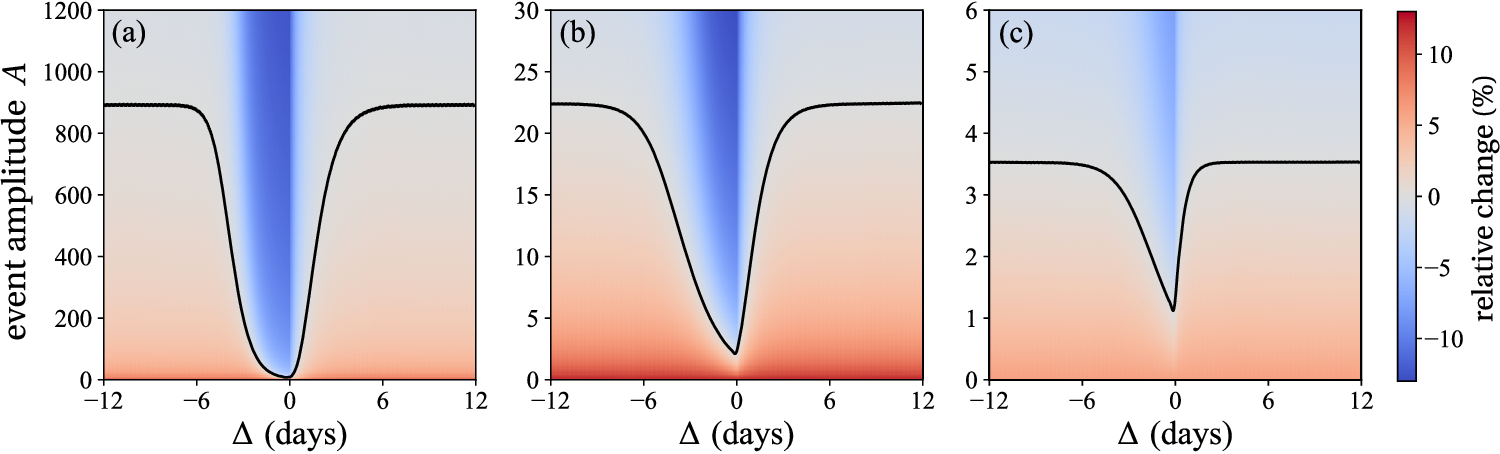}
\caption{Long-term percentage changes in the relapse probability after
  one negative and one positive event relative to the neutral scenario
  of no events occurring.  For $\Delta = \Delta_{\rm ba} = t_2^{\rm a}
  - t_1^{\rm b} > 0$, the stressor $\{ -B_1, t_1^{\rm b} \} $ is
  followed by a positive event $\{A_2, t_2^{\rm a} \} $ where $B_1=
  2$, $t_1^{\mathrm{b}}=12$ days and $A= A_2$ is determined from
  Eq.\,\ref{U_offset_T} at $T = 30$ days.  The color shading
  represents the percentage change of $P(T=30)$ assuming the two
  events have occurred relative to the neutral case.  The black curve
  represents the amplitude of $A_2$ that yields the same relapse
  probability as the neutral case effectively neutralizing the effects
  of the stressor.  For $\Delta = - \Delta_{\rm ab} = t_1^{\rm a} -
  t_2^{\rm b} < 0$ the positive event $\{ A_1, t_1^{\rm a} \} $ is
  followed by a stressor $\{ -B_2, t_2^{\rm b} \} $ where $B_2= 2$,
  $t_2^{\mathrm{b}}=12$ days. Here, the black curve tracks the
  amplitude of a preemptive positive event $A_1$ that would neutralize
  the future $B_2=2$ stressor at $T=30$ days as determined from
  Eq.\,\ref{U_offset_T}.  In panel (a) we set $
  \kappa_\mathrm{a}=2$/day, $\kappa_\mathrm{b}=1$/day; in panel (b) we
  set $\kappa_\mathrm{a}=\kappa_\mathrm{b}=1$/day; and in panel (c) we
  set $\kappa_\mathrm{a}=1$/day, $\kappa_\mathrm{b}=2$/day.  The
  asymmetry between the positive and negative values of $\Delta$
  underlines the importance of event order.  For a given $\Delta =
  \Delta_{\rm ba} > 0$, the value of $A_2$ necessary to balance $B_1 =
  2$ is larger than the value of $A_1$ necessary to balance $B_2 = 2$
  when $\Delta = - \Delta_{\rm ab} = - |\Delta_{\rm ba}| < 0$.  Upon
  comparing the black lines in the three panels for fixed $\Delta =
  \Delta_{\rm ba} >0 $ the value of $A_2$ necessary to neutralize the
  $B_1$ stressor is larger when $\kappa_{\rm a}$ is larger or
  $\kappa_{\rm b}$ is lower, {\textit {i.e.}} under fast processing of
  positive events or longer-lasting stressors.  Similar considerations
  apply when $\Delta = \Delta_{\rm ab} < 0$.}
\label{fig:offset}
\end{figure*}	

We now consider the general case $\kappa_{\rm a} \neq \kappa_{\rm b}$.
In Fig.\,\ref{fig:ABBA_ka_kb} we show results for pairs of events
$\{A_1, t_1^{\rm a} \}$ and $\{-B_2, t_2^{\rm b} \}$ where $A_1 = B_2$
and $\Delta_{\rm {ab}} \equiv t_2^{\rm b} - t_1^{\rm a}$, and for
pairs of events $\{-B_1, t_1^{\rm b} \}$ and $\{A_2, t_2^{\rm a} \}$
where $B_1 = A_2$ and $\Delta_{\rm ba} \equiv t_2^{\rm a} - t_1^{\rm
  b}$.  The long-term relapse probability $P(T)$ is largely
insensitive to the absolute timing of the pair of events, as long as
$T-t_2^\mathrm{a,b}\gg
\kappa_{\mathrm{a}}^{-1},\kappa_{\mathrm{b}}^{-1}$; however, it
strongly depends on the event order and the magnitude of the time lag
$ \Delta_{\rm ab}$ or $\Delta_{\rm ba}$.  For a fixed $ \Delta_{\rm
  ab} = \Delta_{\rm ba}$, $P(T)$ may depend significantly on the order
of events. In general, the larger $\kappa_{\rm a} \Delta_{\rm ab} $
and $\kappa_{\rm b} \Delta_{\rm ba}$, the less sensitive the relapse
probability is to the order of events as shown by comparing the cases
$\Delta_{\rm ab} = \Delta_{\rm ba} = 4$ days with $\Delta_{\rm ab} =
\Delta_{\rm ba} =1$ day in Fig.\,\ref{fig:ABBA_ka_kb}.  Note that
since $\kappa_{\rm a} \neq \kappa_{\rm b}$ it is impossible for the
pairs of events in Fig.~\ref{fig:ABBA_ka_kb} to define the same
integrated mental state $\int_0^{T} M(t') \dd t'$ for any given $T$.

Finally, given a stressor $\{-B_1, t_1^{\rm b} \}$ which is processed
at a rate $\kappa_{\rm b}$ we determine which later event $\{A_2,
t_2^{\rm a} \}$ processed at rate $\kappa_{\rm a}$ will yield the same
relapse probability as the baseline case, where no events occur. To do
this, we note that under the baseline, $U(T) = T$ and that since it is
not possible to define a single event as in Eq.~\ref{dual} for
$\kappa_{\rm a} \neq \kappa_{\rm b}$, we must write $U(T; \Delta_{\rm
  ba})$ in its general form

\begin{equation}
  U(T;\Delta_{\rm ba}) =  t_1^{\rm b}+\int_0^{\Delta_{\rm ba}}
  e^{B_1e^{-\kappa_{\rm b} t}} \dd t + \int_{\Delta_{\rm ba}}^{T-t_1^{\rm b}}\!\!e^{B_1 e^{-\kappa_{\rm b} t' }
          - A_2 e^{-\kappa_{\rm a} (t' - \Delta_{\rm ba})}} \dd t'
\label{U_offset}
\end{equation}
where $\Delta_{\rm ba}= t_2^{\rm a} - t_1^{\rm b}$.  To find the value
of $A_2$ that balances $U(T; \Delta_{\rm ba})$ with the baseline $U(T)
=T$, we must solve
\begin{equation}
  \begin{aligned}
U(T; \Delta_{\rm ba}) = & \, t_1^{\rm b} +\frac{1}{\kappa_{\rm b}} \left[\rm{Ei}(B_1)
  -\rm{Ei} \left(B_1 e^{-\kappa_{\rm b} \Delta_{\rm ba}}\right)\right] \\
\: & \, +  \frac{1}{\kappa_{\rm a}} \left[\rm{Ei}(-A_2)-\rm {Ei}
  \big(-A_2 e^{-\kappa_{\rm a}(T-t_1^{\rm b}-\Delta_{\rm ba})}\big)\right] \\
= & U(T) = T.
\end{aligned}
\label{U_offset_T}
\end{equation}
Under the assumption $\kappa_{\rm a} \Delta_{\rm ba}, \kappa_{\rm b}
\Delta_{\rm ba} \gg 1$, and using the identity
$\lim_{x\rightarrow0}{\rm Ei} (x)=\ln |x|+\gamma$, where $\gamma$ is
the Euler-Mascheroni constant, Eq.\,\ref{U_offset_T} can be simplified
to
\begin{equation}
  \begin{aligned}
\rm{Ei}(-A_2)- \gamma-\ln(A_2) = \frac{\kappa_{\rm a}}{\kappa_{\rm b}}
\big[\ln(B_1) + \gamma - \rm{Ei}(B_1)\big].
  \end{aligned}
\label{offset_simple}
\end{equation}
Eq.\,\ref{offset_simple} yields the value of $A_2$ that neutralizes
the effects of a stressor of amplitude $B_1$ so that at long times the
relapse rate is the same as if neither event occurred.  It is
straightforward to show that upon substituting $B_1 \to -A_1$ and $A_2
\to - B_1$ Eq.\,\ref{offset_simple} still holds when the order of
events is reversed and the positive event occurs prior to the negative
one.  The black curves in Fig.\,\ref{fig:offset} trace the values of
$A = A_2$ that balance a stressor of amplitude $B_1$ given a lag time
$\Delta = \Delta_{\rm ba}> 0 $ for various choices of $\kappa_{\rm a}$
and $\kappa_{\rm b}$.  These results correspond to positive values of
$\Delta$. \textit{Vice-versa}, the amplitude of a protective event $A
= A_1$ that can balance a later stressor of amplitude $B_2$ are shown
for negative values of $\Delta = - \Delta_{\rm ab} < 0 $. The
asymmetry between positive and negative values of $\Delta$ for all
choices of $\kappa_{\rm a}$ and $\kappa_{\rm b}$ implies that a modest
amplitude of the positive event is required to balance a fixed
stressor, regardless of whether the positive event occurs before or
after the stressor. The other color-coded regions in
Fig.\,\ref{fig:offset} show the percent increase (or decrease) of the
relapse probability $P(T)$ compared to the neutral case of no negative
or positive life event, for specific values of $A_1, A_2, B_1, B_2,
t_1^{\rm a}, t_1^{\rm b}, t_2^{\rm a}, t_2^{\rm b}, T, \kappa_{\rm a},
\kappa_{\rm b}$.

\subsubsection{A constant source of
    positivity can offset the random lows of life}
\label{mental1}
\begin{figure}[t!]
  \centering
	\includegraphics[width=0.9\linewidth]{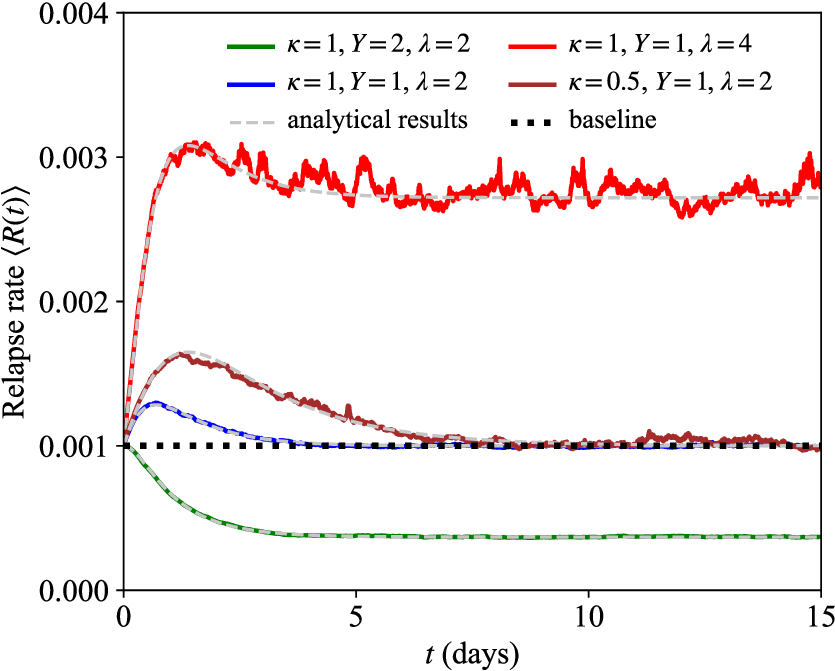}
	\caption{Dynamics of $\langle R(t) \rangle$ averaged over
          100,000 realizations of the Ornstein-Uhlenbeck described by
          Eq.\,\ref{mood} and the analytical expression for $\langle
          R(t) \rangle$ in Eq.\,\ref{ratenoise}. Parameters are
          $M_0=0, R_0 = 10^{-3}$/day.  Values in the legend
          are in units of /day. For $\lambda = 2Y$ the $t \to
          \infty$ relapse rate in Eq. \ref{ratenoiseinfty} $\langle
          R(t \to \infty) \rangle = R_0$ matches the neutral case of
          no exposure to any positive or negative life event.}
\label{relapsestoch}
\end{figure}

\begin{figure*}[t]
  \centering
\includegraphics[width=0.75\linewidth]{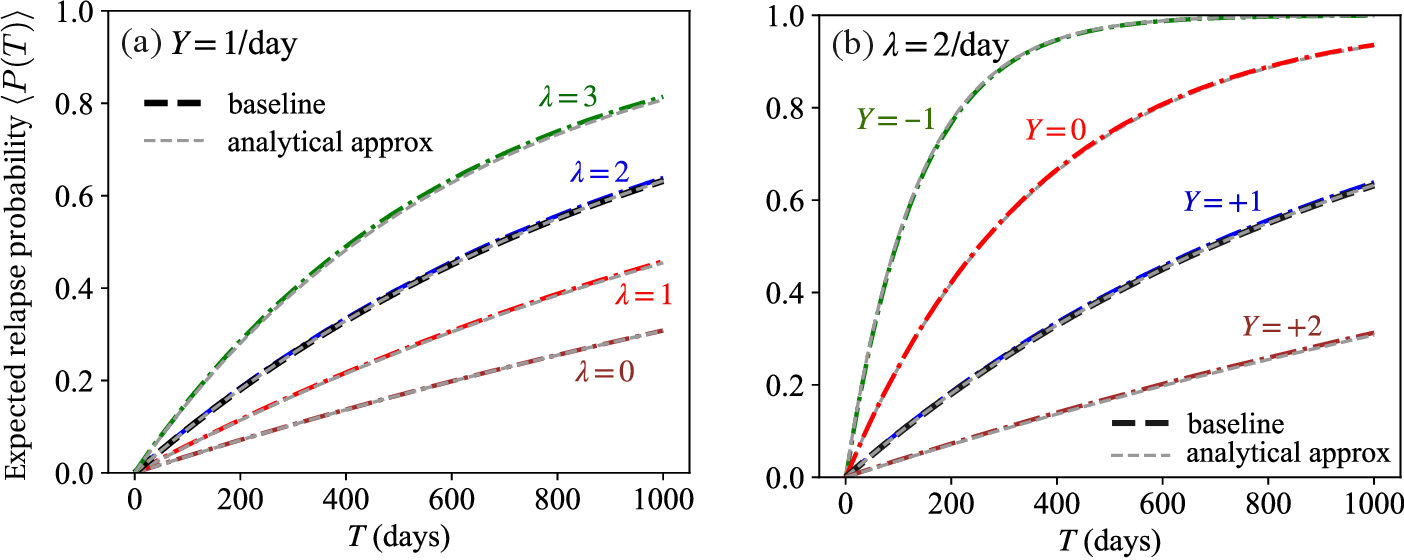}
\caption{Expected relapse probability $\langle P(T) \rangle$ as
  derived from averaging over 5,000 realizations of the
  Ornstein-Uhlenbeck process in Eq.\,\ref{mood} and the analytical
  approximation in Eq.\,\ref{approx} for the baseline $Y= \lambda =0$
  with $R_0 = 10^{-3}, \kappa=1$/day.  Values of $\lambda, Y$
  displayed along the curves are in units of /day. In panel (a) we set
  $Y=1$/day; as $\lambda$ is increased $\langle P(T) \rangle$ also
  increases.  For $\lambda= 2 Y$ results from the baseline are
  recovered.  Small values of $\lambda < 2 Y$ decease $\langle P(T)
  \rangle$ below the baseline, whereas large fluctuations $\lambda > 2
  Y$ increase $\langle P(T) \rangle$ beyond the baseline.  In panel
  (b) we set $\lambda =2$/day and allow $Y$ to be negative,
  representing constant negative experiences that increase $\langle
  P(T) \rangle$.  These results indicate that sustained, large enough
  positive experiences may neutralize a series of random, potentially
  large stressors.}
\label{relapsestoch2}
\end{figure*}

\noindent
We now study the scenario in which there is a constant input $Y$ to
the mental state which may represent a continuous stressor or source
of support. We also assume that positive and negative life events
occur randomly and are processed at rates $\kappa_{\rm a} =
\kappa_{\rm b} = \kappa$. Under these assumptions we 
describe the dynamics for $M = M_{\rm a} + M_{\rm b}$ as

\begin{equation}
	\label{mood}
	\frac{\dd M}{\dd t} = - \kappa M + Y + \xi(t), 
\end{equation}
where $\xi(t)$ is a Gaussian white noise term that represents the
random, positive and negative, life events. We set the general initial
mental state value $M(t=0) = M_0$ and write the mean and correlation
function of $\xi(t)$ as
\begin{equation}
  \begin{aligned}
	\langle \xi (t) \rangle =& 0 \\
	\langle \xi (t) \xi (t') \rangle =&  2 \lambda \delta(t-t'). 
        \label{correlate}
        \end{aligned}
  \end{equation}
Eqs.~\ref{correlate} define an Ornstein-Uhlenbeck stochastic process,
a classic paradigm in statistical mechanics
\cite{Uhlenbeck1930,Gardiner1985,Risken1989,Zeng2020}.  The relapse
rate of the neutral case, where there are no life events or continuous
inputs to perturb an individual's mental state and $Y = \xi(t) = 0$,
is given by
\begin{equation}
	\label{cave}
	R_{\rm d} (t) = R_0 e^{- M_0 e^{-\kappa t}}. 
\end{equation}
If $Y \neq 0$ and $\xi(t) \neq 0$, Eq.~\ref{mood} can be solved as
\begin{equation}
  \begin{aligned}
	M(t) = & \mu_{\rm m}(t) +  \int_{0}^t\! e^{-\kappa (t-t')} \xi(t') \dd t',\\
	\mu_{\rm m}(t) \equiv &  \frac{Y}{\kappa}
        + \Big(M_0 - \frac{Y}{\kappa} \Big) e^{-\kappa t}. 
 	\label{avg}
  \end{aligned}
  \end{equation}
Associated with Eqs.~\ref{mood} and \ref{correlate} is a Fokker-Planck
equation governing the dynamics of the probability density function
$P_{\rm m}(M,t)$ \cite{Risken1989, Gardiner1985}
\begin{equation}
	\label{FP}
	\frac{\partial P_{\rm m}(M,t) }{\partial t}
        =  \frac{\partial}{\partial M}
        \Big((\kappa M - Y) P_{\rm m}\Big)
        + \lambda \frac{\partial^2 P_{\rm m}}{\partial M^2}.
\end{equation}
The solution to Eq.~\ref{FP} and the
initial condition $P(M, t=0) = \delta(M - M_0)$
is
\begin{equation}
  \begin{aligned}
    P_{\rm m}(M,t)  = & \frac{1}{\sqrt{2 \pi  \sigma_{\rm m}^2 (t)}}
    e^{-\frac{(M - \mu_M(t))^2} {2 \sigma_{\rm m}^2 (t)}} \\
	\sigma^2_{\rm m}(t) \equiv & \frac{\lambda}{k} (1 - e ^{-2 kt}).
\end{aligned}
  \label{distribution}
\end{equation}
In the limit $t \to \infty$, Eq.\,\ref{distribution} defines a steady
state Gaussian distribution centered at $Y/\kappa$ with variance
$\lambda/\kappa$.  The expected time-dependent relapse rate is given
by

\begin{equation}
  \langle R(t) \rangle = R_0 \langle e^{-M (t)} \rangle
  = R_0\!\int_{-\infty}^{\infty}\!\!\!e^{-M} P(M,t) \dd M 
\end{equation}
which is explicitly expressed as 
\begin{subequations}
  \begin{align}
\langle R(t) \rangle = & R_0
	\exp \Big[ - \mu_{\rm m}(t)+\frac{1}{2}\sigma_{\rm m}^2(t)\Big],\label{ratenoise}\\
	\langle R(t \to \infty) \rangle  = & R_0 \exp 
	\Big[\frac{(\lambda - 2Y)}{2 \kappa}\Big].\label{ratenoiseinfty}
        \end{align}
\end{subequations}
The relapse rate in Eq.~\ref{ratenoiseinfty} can be compared with the
equivalent expression obtained in the absence of noise, {\textit
  {i.e.}} for $Y \neq 0$ and $\xi(t)=\lambda = 0$
\begin{equation}
	\label{ratio0}
	\frac{\langle R(t \to \infty) \rangle}{R(t \to \infty;\xi(t)=0)}= 
	e^{ \lambda / 2 \kappa} > 1.
\end{equation}
Since the above $\lambda$-independent ratio is always larger than one,
Eq.\,\ref{ratio0} implies that unbiased noise, where positive and
negative life events are equally likely in frequency and magnitude,
results in a larger relapse rate than the noise-free case, regardless
of the sign of $Y$.  Mathematically, this result stems from the
asymmetry in $R_0 e^{-M(t)}$ where the increase due to a stressor is
much larger than the decrease following a positive fluctuation
of similar amplitude, consistent with the brain's negativity bias
\cite{Ito1998}.  We can also compare the general case $Y \neq 0$ and
$\xi(t) \neq 0$ with the neutral case $Y = \xi(t) = 0$ yielding
\begin{equation}
  \frac{\langle R(t \to \infty) \rangle}{R_{\rm d}(t \to \infty)} =
\exp \Big[ \frac{(\lambda - 2 Y )}{2 \kappa} \Big].
\label{ratio}
\end{equation}
The values of $Y, \lambda $ in Eq.\,\ref{ratio} can be tuned so that
the driving term and the noise balance each other. Specifically, for
the long term expected relapse rate in the presence of noise to be
less than the relapse rate in the absence of any external input, the
constant input $Y$ must obey $Y > \lambda/2$. Given the form for
$P(T)$ in Eq.\,\ref{survive}, we can write the expected relapse
probability as
\begin{equation}
\label{relaprob}
	\langle P(T) \rangle =  
	1-\left\langle \exp \Big[-R_0 \int_0^T e^{-M(t)} \dd t \Big]
        \right\rangle
\end{equation}
and approximate it in the $\kappa T  \gg 1$ limit by
\begin{equation}
\begin{aligned}
  \langle P(T) \rangle & \approx  1-\exp
\Big[-R_0 \int_0^T \langle e^{-M(t)} \rangle \dd t \Big]\\
\: & \approx  1-\exp \Big[- e^ {\frac{\lambda - 2 Y }{2 \kappa}}R_0 T \Big].
        \end{aligned}
	\label{approx}
\end{equation}
In the Appendix we discuss Eq.\,\ref{relaprob} and cases where the
approximation in Eq.\,\ref{approx} fails, namely in the $\kappa \to 0$
limit.  Fig.\,\ref{relapsestoch} shows the expected relapse rate
$\langle R(t) \rangle$ as $\kappa, Y, \lambda$ are varied.  Results
derived from 100,000 simulations of the stochastic process in
Eq.\,\ref{mood} are compared with predictions from the analytical
result (Eq.\,\ref{ratenoise}). We also show the relapse rate for the
baseline given by $R_{\rm d}(t) = R_0$. The expected relapse rate
increases with the noise amplitude $\lambda$ and decreases with the
magnitude of the positive experience $Y$ and with the processing rate
$\kappa$.  Lower $\kappa$ values imply longer processing times for all
events; however, since the asymmetry in $R(t)$ assigns more weight to
negative occurrences, a larger likelihood of relapse is observed as
$\kappa \to 0$.  The corresponding values of the expected relapse
probability $\langle P(T) \rangle$ obtained from simulations and from
the analytical approximation in Eq.\,\ref{approx} are shown in
Fig.\,\ref{relapsestoch2}.

\begin{figure*}[t]
  \centering
 \includegraphics[width=0.75\linewidth]{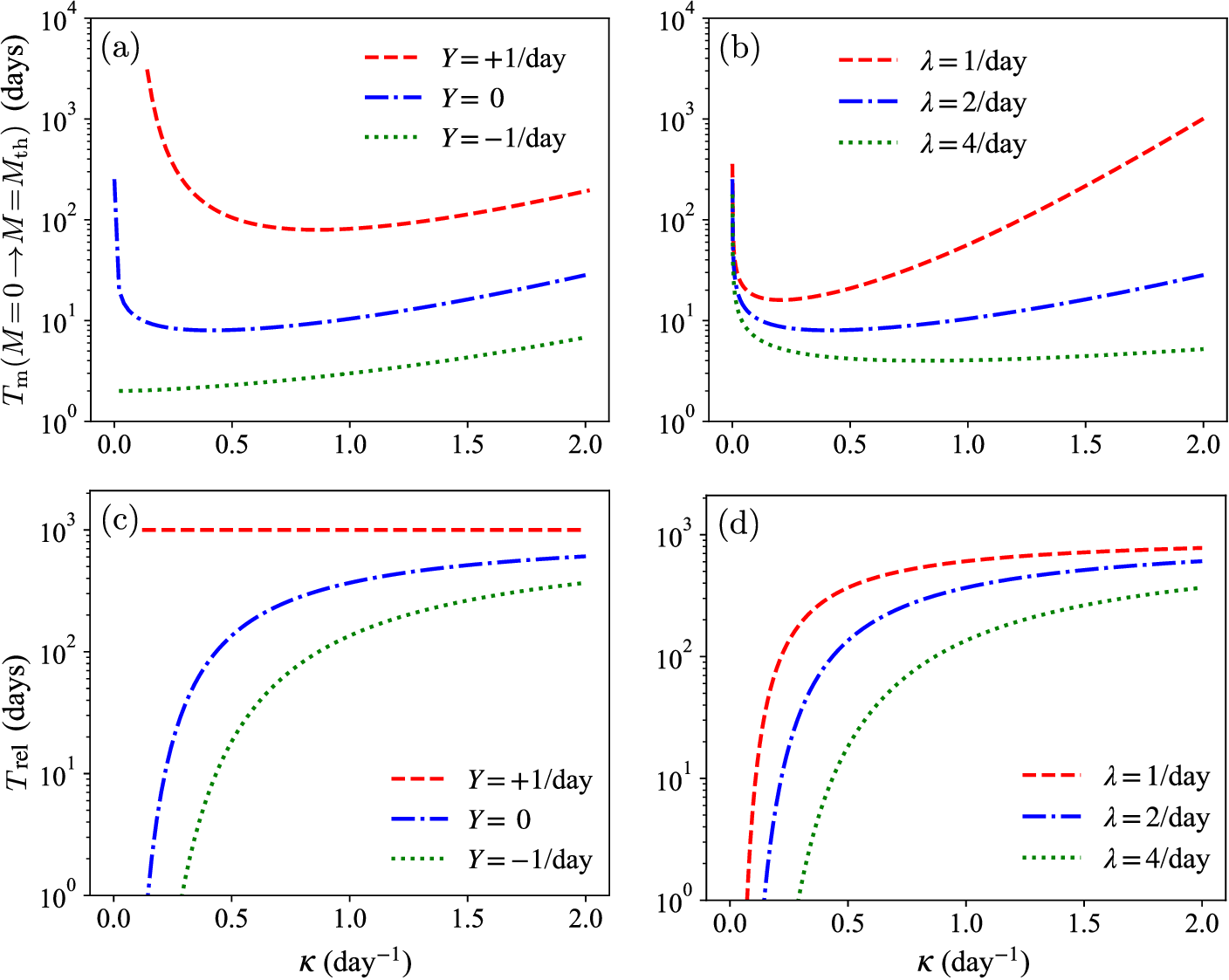}
 \caption{Top row: Mean first passage time $T_{\rm m} (M = 0\to
   M=M_{\rm th})$ for an initial mental state $M=0$ to reach the
   negative threshold $M_{\rm th} = -2$ as a function of the decay
   rate $\kappa$.  Results follow from the random process in
   Eq.\,\ref{mood} and the analytical form in Eq.\,\ref{MFPT}. For
   large $\kappa$, each random fluctuation dissipates quickly so that
   the effects of multiple inputs to the mental state do not
   accumulate appreciably, hence $T_{\rm m}(M)$ increases with
   $\kappa$.  Increasing $\kappa$ for small values of $\kappa$ will
   reduce the likelihood of positive values of the mental state thus
   shortening the time to reach $M_{\rm th}$ as discussed in the text.
   These two trends lead to the observed non-monotonic behavior, which
   is most pronounced for large $Y, \lambda$.  In (a) we set $\lambda
   = 2$/day and vary the constant input $Y$; in (b) we set $Y = 0$ and
   vary the noise amplitude $\lambda$.  Bottom row: Mean first passage
   time $T_{\rm rel}$ to relapse computed from the approximation in Eq.~\ref{MFPT_REL} and
   shown as a function of the decay rate $\kappa$. In panel (c) we set
   $\lambda = 2$/day and vary the constant input $Y$; in panel (d) we
   set $Y = 0$ and vary the noise amplitude $\lambda$.}
	\label{fig:MFPT}
	\end{figure*}

Finally, we evaluate the first passage statistics to a given mental
state $M_{\rm th} <0$. Although the threshold level value can be
arbitrary, we set it to be negative to represent a critically unhappy
mental state.  The dynamics of the mean first passage time $T_{\rm
  m}(M)$ to reach $M_{\rm th}$ starting from a given mental state
$M > M_{\rm th}$ can be derived from the backward Kolmogorov
equation associated with Eq.\,\ref{FP}

\begin{equation}
	\label{BFP}
	\lambda \frac{\dd^2 T_{\rm m}}{\dd M^2}
        - (\kappa M - Y) \frac{\dd T_{\rm m}} {\dd M} = - 1
\end{equation}
along with absorbing boundary conditions $T_{\rm}\, (M_{\rm th}) =
0$ and $T_{\rm m}\, (M \to \infty) = 0$. Equation \ref{BFP} can be
solved using standard methods to yield

\begin{equation}
\label{MFPT}
T_{\rm m}(M)  =\frac{ \sqrt {\pi}}{\kappa}
\int_{ \sqrt{\frac{\kappa}{2 \lambda}}(M_{\rm th} - Y/\kappa)}
^{\sqrt{\frac{\kappa}{2 \lambda}} (M - Y/\kappa)}\! e^{z^2} {\rm{erfc}} (z) \dd z
\end{equation}
where ${\rm erfc}(x)$ is the complementary error function ${\rm
  erfc}(x) = 1 - {\rm{erf}}(x)$.  Since the argument of the integrand
is a positive, decreasing function of $z$, Eq.\,\ref{MFPT} implies
that $T_{\rm m}(M) $ is increasing in $Y, M$ and decreasing in
$\lambda, M_{\rm th}$. This is to be expected, given that large values
of $Y, M$ tend to shift the mental state away from the lower negative
threshold $M_{\rm th}$, and given that large $\lambda$ values lead to
larger fluctuations that are more likely to reach the negative $M_{\rm
  th}$. However, $T_{\rm m}(M)$ is non-monotonic with $\kappa$ as
shown in the top row of Fig.\,\ref{fig:MFPT}. Here we plot $T_{\rm m} (M=0)$ as
derived from Eq.\,\ref{MFPT} and denote it as $T_{\rm m}(M=0 \to M =
M_{\rm th})$ for clarity.  Note that $T_{\rm m}(M=0)$ decreases with
increasing $\kappa$ at low values of $\kappa$ and increases with
$\kappa$ at large $\kappa$. To understand this non-monotonicity, note
that the mental state $M$ will be restored towards $Y/\kappa$ within a
time frame $\kappa^{-1}$ after any random fluctuation. As $\kappa \to
\infty$, this implies that after each random event, $M \to 0$ very
quickly and that cumulative effects of multiple past random events
will be very limited.  As a result, increasing $\kappa$ when $\kappa$
is already large will make reaching $M_{\rm th}$ less likely and thus,
the mean first passage time will increase.

On the other hand, when $\kappa \approx 0$, $M$ will grow
approximately linearly away from $M_{\rm th}$ driven by the $Y$ input
while being subject to noise. Whatever the initial value of $M$, for
large enough values of $Y$ and $\lambda$, $M$ will most likely become
positive before one, or a sequence of, negative random events will
lead it close to $M_{\rm th}$. Increasing $\kappa > 0$ will now lessen
any positive values of $M$, and thus accelerate reaching the $M_{\rm
  th}$ threshold.  Any negative values of $M$ will instead increase as
$\kappa$ is increased, and this will cause a delay in hitting $M_{\rm
  th}$.  However, given the negative value of the threshold,
excursions in the $M >0$ space will be much longer than those in the
$0 < M < M_{\rm th}$ space, so that acceleration in reaching $M_{\rm
  th}$ will dominate.  As a result, increasing $\kappa$ when $\kappa$
is small will lead to a shorter mean first passage time. This effect
will be more pronounced for large values of $Y, \lambda$ as the
excursion to, and permanence in, the $M > 0$ space will be more
sustained in these cases.  When $Y, \lambda$ are small and especially
if the initial $M$ value is negative, decreases in $T_{\rm m}(M)$ as
$\kappa$ is increased for small values of $\kappa$ will be much less
evident, as excursions to the $M>0$ plane will be rare. An alternate
representation of Eq.\,\ref{MFPT} is given by expanding the integrand
via a Taylor series and evaluating the integral, leading to

\begin{equation}
  \begin{aligned}
    T_{\rm m} (M)  = & \frac{\pi}{2 \kappa}{\rm erfi}(z_{\rm in})
    - \frac{z_{\rm in}^2}{\kappa}
{_2}F_2 \Big( 1,1; \frac{3}{2}, 2; z_{\rm in}^2\Big) \\
\: & - \frac{\pi}{2 \kappa}{\rm erfi}(z_{\rm eg})
+ \frac{z_{\rm eg}^2}{\kappa}
{_2}F_2 \Big( 1,1; \frac{3}{2}, 2; z_{\rm eg}^2 \Big) 
\end{aligned}
  \label{MFPT2}
\end{equation}
\noindent
where ${_2}F_2$ is the generalized hypergeometric function and 
\begin{equation}
  \begin{aligned}
    z_{\rm in} = \sqrt{\frac{\kappa} { 2 \lambda}}
    \Big(M - \frac{Y}{\kappa} \Big),
    \,\,\, z_{\rm eg} =  \sqrt{\frac{\kappa}{2 \lambda}}
    \Big(M_{\rm th} - \frac{Y}{\kappa} \Big). 
  \end{aligned}
  \end{equation}
  
Eqs.\,\ref{MFPT} and \ref{MFPT2} are the mean time $T_{\rm m}(M)$ for
an initial mental state $M$ to first reach the threshold $M_{\rm th}$;
one may similarly determine the mean first time to relapse $T_{\rm
  {rel}}$ using Eq.\,\ref{survive} 
\begin{equation}
  \begin{aligned}
   T_{\rm {rel}} = \int_0^{\infty} \langle S(T) \rangle \dd T 
   \approx \frac{e^{\frac{2 Y - \lambda}{2 \kappa}}} {R_0}.
  \end{aligned}
\label{MFPT_REL}
  \end{equation}
The last relationship arises from Eq.\,\ref{approx} and is
valid for $\kappa T \gg 1$. 
We show $T_{\rm {rel}}$ as a function of $\kappa$ in the bottom row of
Fig.\,\ref{fig:MFPT}, using the same parameters chosen for $T_{\rm
  {m}}(M)$. The mean first time to relapse is an increasing function
of $Y$ and a decreasing function of $\lambda$, indicating that
positive continuous inputs and exposure to
relatively small noise amplitudes can act as protective factors. These
trends are consistent with those observed for $T_{\rm {m}}(M)$.

\subsection{The presence of cues}	

In this section, we study how cues affect the mental state and the
likelihood of relapse.  According to Eqs.~\ref{rate} and \ref{wpeak}
sensory cues are mathematically represented as a stressor of fixed
amplitude $w_{\rm peak}$ occurring at times $t_{\ell}^{\rm c}$.  This
description is consistent with psychiatric studies that have
identified overlaps in the neural circuits that process stress and
drug-related cues and that have found that both lead to cravings and
heightened susceptibility to relapse \cite{Fox2005, Sinha2007}.  As
mentioned earlier, we assume that cues always bring back memories of
the first high so that the amplitude $w_{\rm peak}$ is fixed. As a
result, findings illustrated in Sections \ref{nocues1} and
\ref{nocues4} still hold upon substituting $B_{i} \to w_{\rm peak}$
for all $i$ and $\kappa_{\rm b} \to \kappa_{\rm c}$.  In particular,
relapse is less likely if a positive experience occurs immediately
after being exposed to a drug-related cue and one can still utilize
the results shown in Fig.~\ref{fig:offset} to determine the magnitude
and timing of the positive experience necessary to balance exposure to
a cue.  Values of $\kappa_{\rm c}$ will be chosen such that
$\kappa_{\rm c} > \kappa_{\rm b}$, as we expect the time to process a
drug-related cue to be less than the time to overcome a stressor.

Consider the case in which an individual is randomly exposed, through
a Poisson process with rate $\lambda_{\rm c}$, to cues that elicit the
memory of the first high. The individual thus experiences, with
probability $P_{\rm c}(n_{\rm c},t) = (\lambda_{\rm c} t)^{n_{\rm c}}
e^{-\lambda_{\rm c} t}/n_{\rm c}!$, $n_{\rm c}$ cues within a time
interval $t$. The mean time between successive cues is $1/\lambda_{\rm
  c}$.  Given the equivalence between cues and stressors, these
results may also be interpreted as the response to an identical
stressor presented at random, Poisson-distributed times.  The dynamics
of the cue-induced motivation is thus
\begin{equation}
\label{poisson1}
	\frac{\mathrm{d}C}{\dd t}=-\kappa_{\rm c} C + w_{\rm peak} 
	\sum_{\ell = 1; t \geq t^{\rm c}_{\ell}}^{n_{\rm c}}  \delta(t-t^{\rm c}_{\ell}),
\end{equation}
where $n_{\rm c}$ are the Poisson-distributed number of events that
have occurred until time $t$.  We also assume that there is a
counteracting source of support $Y > 0 $ to the mental state so that
\begin{equation}
\label{counterm}
M(t) = M_{\rm a}(t) = \frac{Y}{\kappa_{\rm a}} \big(1 - e^{- \kappa_{\rm a} t}\big).
\end{equation}
We solve Eq.~\ref{poisson1} with the initial condition $C(t=0) =
0$. The resulting expression for $C(t)$ can be used to write the
relapse rate $R(t)$ in Eq.~\ref{rate} using the mental state $M(t) =
M_{\rm a}(t)$ given in Eq.~\ref{counterm} and under the assumption
that drugs are fully available $I(t) =1$. We find

\begin{equation}
\label{cues}
	\frac{R(t)}{R_0} =  
	e^{-\frac{Y}{\kappa_{\rm a}}\left(1-e^{-\kappa_{\rm a} t}\right)}
	 \!\!\prod_{\ell =1; t \geq t^{\rm c}_{\ell}}^{n_{\rm c}} 
	 \!\!e^{w_{\rm peak}e^{-\kappa_{\rm c} \left(t-t_{\ell}^{\rm c}\right)}.}
\end{equation}
Using well known properties of the Poisson process shown in the
Appendix we estimate the expected value of the relapse rate for any
number of events occurring within time $t$ as
\begin{figure*}[t!]
  \centering
 \includegraphics[width=0.9\linewidth]{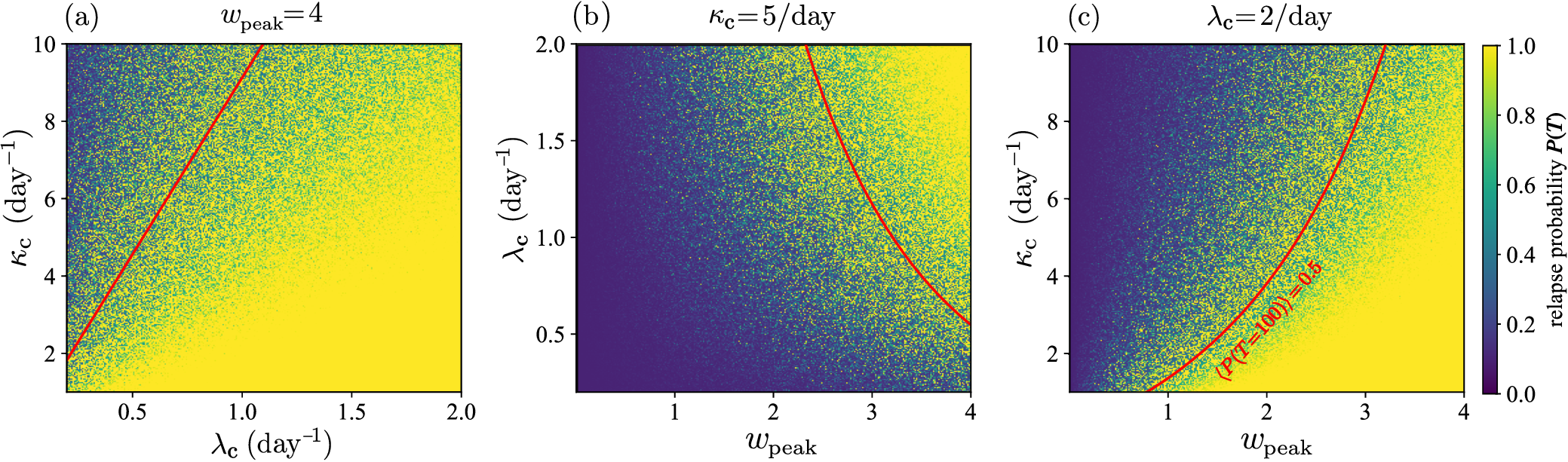}
\caption{Relapse probability $P(T=100)$ following exposure to a
  single, randomly drawn sequence of sensory cues occurring at times
  generated according to a Poisson process of rate
  $\lambda_\mathrm{c}$.  The corresponding relapse rate is evaluated
  through Eq.\,\ref{cues} and the relapse probability at $T=100$ days,
  $P(T=100)$, is evaluated using Eqs.\,\ref{survive}. The red curves
  depict the analytical approximation obtained for $\langle
  P(T=100)\rangle=0.5$ calculated using Eqs.\,\ref{ln_R_infty} and
  \ref{approxcue}. We do not include any positive, continuous form of
  support so that $Y=0$ and results are independent of the processing
  rate $\kappa_{\rm a}$.  In panel (a) we fix $w_{\rm peak}=4$; in
  panel (b) we fix $\kappa_{\mathrm{c}}=5$/day; in panel (c) we fix
  $\lambda_{\mathrm{c}}=2$/day. Collectively, the panels show that the
  likelihood of relapse increases with increases in the amplitude
  $w_\mathrm{peak}$, the frequency of cue occurrences
  $\lambda_\mathrm{c}$, or the typical cue processing time
  $1/\kappa_{\mathrm{c}}$.}
  \label{fig:cuepoisson}
\end{figure*}

\begin{equation}
  \begin{aligned}
  \label{mascheroni}
	\left\langle{\frac{R(t)}{R_0}}\right\rangle = & 
 	\exp\left({- \lambda_{\rm c} t
          -\frac{Y}{\kappa_{\rm a}}(1-e^{-\kappa_{\rm a} t})}\right) \\
        \: & \,\,\,\times \exp\Big[\lambda_{\rm c}\int_0^t
          e^{w_{\rm peak} e^{-\kappa_{\rm c}\left(t-t'\right)}}
          \dd t' \Big].
  \end{aligned}
  \end{equation}
In the $\kappa_{\rm a} t, \kappa_{\rm c} t \gg 1$ limit
Eq.\,\ref{mascheroni} can be approximated by
\begin{equation}
	\ln \Big\langle{\frac{R(t)}{R_0}}\Big\rangle \approx
	-\displaystyle{\frac{Y}{\kappa_{\rm a}}
          +\frac{\lambda_{\rm c}}{\kappa_{\rm c}}} \big({\rm Ei}
	(w_{\rm peak})-\ln(w_{\rm peak})-\gamma\big),
  \label{ln_R_infty}
\end{equation}
where $\gamma$ is the Euler-Mascheroni constant.  We use the
approximation in Eq.~\ref{ln_R_infty} to evaluate the
relapse rate $R(t)$ in Eq.~\ref{survive}, leading to an approximate
expression for $\langle P(T) \rangle$, valid for $\kappa_{\rm a} T,
\kappa_{\rm c} T \gg 1$

\begin{equation}
\label{approxcue}
\langle P(T) \rangle = 1 - \langle e^{- \int_0^T R(t) {\rm d} t} \rangle
\approx 1 - e^{-  R_0 T e^{Q_{\rm c}}},
\end{equation}

\noindent
where
\begin{equation}
  Q_{\rm c} \equiv
  -\displaystyle{\frac{Y}{\kappa_{\rm a}}+\frac{\lambda_{\rm c}}{\kappa_{\rm c}}}
  \big({\rm Ei}(w_{\rm peak})-\ln(w_{\rm peak})-\gamma\big).
\end{equation}

In Fig.~\ref{fig:cuepoisson}, we evaluate the probability of relapse
at given values $\kappa_{\rm c}, \lambda_{\rm c}, w_{\rm peak}$. We do
this by first drawing a sequence of cues that occur
according to a Poisson process of rate $\lambda_{\rm c}$.  Then, using
this specific sequence of cues, we compute the relapse rate given in
Eq.~\ref{cues}. Finally, we determine the survival probability and
the relapse probability at a fixed time $T = 100$ days using
Eqs.~\ref{survive}. The results in Fig.~\ref{fig:cuepoisson} are
obtained under the assumption of no positive input, $Y = 0$.  For
comparison we use the analytical approximation given in
Eqs.\,\ref{ln_R_infty} and \ref{approxcue} to find the parameter curve
corresponding to $\langle P(T=100) \rangle = 0.5$.

In the baseline case, where there are no cues or external stimuli,
$R(t)=R_0$, hence the positive support $Y$ will balance or alleviate
the cue-induced drive to take drugs only if it satisfies
\begin{equation}
	Y \ge \frac{\lambda_{\rm c} \kappa_{\rm a}}{\kappa_{\rm c}}
	\big({\rm Ei}(w_{\rm peak})-\ln(w_{\rm peak})-\gamma\big).
\end{equation}
Thus, to be effective, the counterbalancing source of support $Y$ must
increase with the intensity of the memory of the first high $w_{\rm
  peak}$, the rate of exposure to the cues $\lambda_{\rm c}$, the
duration of the cue $1 / \kappa_{\rm c}$ and the processing rate of
the positive events $\kappa_{\rm a}$.

\section{Conclusions}
\label{conclude}
We presented a mathematical model for the probability of relapse in
drug addiction.  Our model incorporates dynamics that reflect
psychiatric concepts such as the positive activation, negative
activation (PA/NA) model and the peak-end rule.  

Addiction research, like other studies that focus on learning, memory,
rewards and synaptic plasticity, relies on neuroimaging methods to
understand how the brain and its neurocircuitry adapt to short- or
long-term drug use and ensuing behavioral changes. It is well
documented that drug users and former users display dysregulation in
their brain reward system, heightened reactivity to drug-related cues
and stressors, less inhibitory self-control, and a tendency to engage
in compulsive behaviors.  It is also well established that the process
of physical detoxification is a relatively short one, but cravings and
relapse can occur even long after cessation of drug use. Relapse is
often triggered by exposure to stressors or drug-related cues that the
former user is unable to manage.  Among the biggest limitations of
neuroimaging studies and clinical trials are the need to control for
individual predispositions and external circumstances, high costs,
difficulties in recruiting volunteers with substance use disorder
especially in longitudinal studies.

Given the complexities of addiction and the practical limitations in
obtaining comprehensive data, simple and analytically tractable
mathematical models may be helpful to understand how the brain
responds to drugs and their absence. Decision-making and many
psychiatric disorders, including addiction, have been described using
quantitative mathematical models in recent years
\cite{Gauld2023,Trofymchuk2023,Cheng2020,Hauser2022,Kim2016,Harris2015}.
In our work, we considered the response of the brain to a series of
inputs representing positive and negative events, and how their
amplitude, timing and ordering affect the likelihood that a person in
recovery will use again. By construction, and mirroring the PA/NA
model, negative events increase the likelihood of relapse more than
positive ones of the same magnitude.  We find that clustering positive
or negative events is generally detrimental.  For a fixed, mental
state activity integrated over a fixed time frame and imparted by an
arbitrary number of negative (or positive) events, the best way of
distributing these events is through a continuum of moderate
negativity (or positivity), rather than as a large jolt of catastrophe
(or happiness) occurring at all once.  On the other hand, once an
individual is exposed to a stressor, a positive event occurring
immediately afterward can act as a protective factor. We also found
that a constant source of positive input can balance the negativity
arising from a series of random events that may include large
stressors.  Since the mathematical representation of sensory cues in
our work is akin to that of stressors, the above considerations remain
valid for exposure to drug-related cues.

The effect of different environments and user profiles may be studied
by tuning relevant parameters. By changing $\kappa_{\rm a},
\kappa_{\rm b}, \kappa_{\rm c}$ and $w_{\rm peak}$ we can represent
users who respond differently to life experiences and whose memories
of the ``first high'' vary in intensity. Similarly, the amplitudes
$A_i, B_j$, the Gaussian noise $\lambda$ and the Poisson parameter
$\lambda_{\rm c}$ can represent different risk levels in the social
environment of the recovering addict. Finally, although developed in
the context of relapse, our model can be used to also study the
driving and protective factors that lead a non-user to try drugs for
the first time. In this case, $C(t) =0$ as there are no sensory cues
related to past use, but $M(t)$ can represent external stimuli that
induce an individual to use drugs for the first time.

How do we translate these findings into practice? How to experience
continuous positivity? Certainly, it is important to seek out
positive, fulfilling experiences, embodied by the $A_i$ events
discussed in this work. However, the continuous sources of positivity
we introduced, such as the green curves in Fig.\,\ref{fig:small
  events} and the $Y$ term in Eq.\,\ref{mood}, represent inputs to the
mental state.  One may interpret these inputs as arising not just from
actual events, but also as imparted from a positive attitude towards
life, for example through support from family and friends, finding
satisfaction in one's work, hobbies and social life. A positive
attitude can also be developed through cognitive behavioral therapy,
individual or group counseling or psychotherapy which are known to be
effective in helping manage life's challenges without recourse to
drugs.
%
%
\begin{table*}[t!]
\label{tablepar}
\renewcommand\arraystretch{1.5}
\centering
\begin{tabular}{lll}
		\toprule[2pt]
		\textbf{SYMBOL~~~}&\textbf{QUANTITY}&\textbf{RANGE}\\
		\midrule[1pt]
		$M_{\rm a}$ & positive activity of the mental state & $\sim 10$\\
		$M_{\rm b} $ & negative activity of the mental state & $\sim -10$\\
		$C$ & mental response to cues & $\sim 1$\\
		$\kappa_{\rm a},~\kappa_{\rm b}$ & equilibration values of
                the mental state processing rates & $\sim 1$/day\\
		$\kappa_{\rm a,0},~\kappa_{\rm b, 0}$ & onset values of the mental
                state processing rates & $0.6 \sim 0.8 \kappa_{\rm{a,b}}$
                \cite{volkow2001loss}\\
		$\gamma_{\rm a},~\gamma_{\rm b}$ & recovery rates for $\kappa_{\rm a}(t)$
                and $\kappa_{\rm b}(t)$ & $0.002\sim 0.02$/day
                \cite{volkow2001loss}\\
		$\kappa_{\rm c}$ & processing rates of drug-related cues and
                memories & $\sim 10$/day\\
		$R_0$ & rate of relapse in the neutral mental state (without inputs) &
                $\sim 10^{-3}$/day\\
		$A_i$ & intensity of positive life event \textit{i} & $\sim 1$\\
		$B_j$ & intensity of negative life event \textit{j} & $\sim 1$\\
		$w_{\mathrm{peak}}$& intensity of most pleasurable drug-taking reward
                response & $\sim 1$\\
		$t_i^\mathrm{a}$ & time of occurrence of positive life event
                \textit{i} & $\sim$ day\\
		$t_j^\mathrm{b}$ & time of occurrence of negative life event
                \textit{j} & $\sim$ day\\
		$t_\ell^\mathrm{q}$ & time of occurrence of cue $\ell$&$\sim$ day\\
		$\lambda_q$ & Poisson process rate for the occurrence of cues &
                $\sim 1$/day\\
		$Y$ & continuous input to the mental state&\\
		$\kappa$  & mental state processing rate for
                $\kappa_{\rm a} = \kappa_{\rm b}$ \\
		$\lambda$ & Gaussian noise intensity in the OU process
                for the mental state &\\
		\bottomrule[2pt]		
	\end{tabular}
	\caption{Relevant quantities and parameter ranges for the
          relapse models presented in Eqs.\,\ref{rate}, \ref{mooda},
          \ref{moodb} (distinct mental state neurocircuitry for few
          positive and negative events, no cues); Eqs.\,\ref{rate},
          \ref{mood}, (common mental state neurocircuitry for random,
          Gaussian distributed positive and negative events, no cues);
          Eqs.\,\ref{rate}, \ref{wpeak}, \ref{poisson1},
          \ref{counterm} (distinct mental state neurocircuitry for
          uniform, positive events, subject to Poisson distributed
          drug-related cues) and time-dependent forms for the
          processing rates.}
          \label{tablepar}
\end{table*}

\section{Acknowledgement}

M.R.D. and T. C.  acknowledge support from the Army Research Office
through grant W911NF-18-1-0345.

\section{Appendix}
\appendix

\section{Time-dependent processing rates $\kappa_{\rm a}(t)$ and $\kappa_{\rm b}(t)$ }

Time-dependent processing rates $\kappa_{\rm a}(t)$, $\kappa_{\rm b}
(t)$ are included in our mathematical representations of the PA/NA
model in Eqs.\,\ref{mooda} and \ref{moodb} to allow for neuroadaptive
changes after cessation of drug use. Assuming that at the beginning of
the recovery phase at $t=0$ there are no negative or positive affects
so that $M_{\rm a}(t=0) = M_{\rm b}(t=0) =0$, the general solution is
\begin{subequations}
  \begin{align}
	M_{\rm a}(t) = &  \sum_{i, t \geq t_{i}^{\rm a}}
        A_i e^{- \int_{t_{i}^{\rm a}}^{t} \kappa_{\rm a}(s') {\rm d} s'}, \label{Ma_outcome}  \\
	M_{\rm b}(t)  = & - \sum_{j, t \geq t_{j}^{\rm b}}  B_j e^{- \int_{t_j^{\rm b}}^{t}
          \kappa_{\rm b}(s') {\rm d} s'}	\label{Mb_outcome}
        \end{align}
\end{subequations}
Drug addiction is known to attenuate the pleasure stemming from
positive stimuli, leading to anhedonia \cite{Koob2022b}.  It is thus
reasonable to assume that after cessation of drug use, positive
experiences will return to being more pleasurable.  This is supported
by experimental evidence that dopamine transporter loss in former
methamphetamine users can recover after a sufficiently long period of
abstinence \cite{volkow2001loss}.  Plausible forms for $\kappa_{\rm
  a}(t)$ include monotonically decreasing functions that start at
$\kappa_{\rm a}(t=0) = \kappa_{\rm a,0}$ and that descend towards the
standard value at full recovery $\kappa_{\rm a} (t \to \infty) =
\kappa_{\rm a}^* < \kappa_{\rm a,0} $.  Within this scenario, positive
events occurring well into the recovery phase elevate one's mental
state for a longer period compared to positive events occurring at the
onset of the recovery phase. Similarly, since drug abuse is known to
exacerbate negative emotional distress \cite{Koob2020, Koob2020b,
  Koob2021, Cheetham2010}, we can assume that $\kappa_{\rm b}(t)$ is a
monotonically increasing function with $\kappa_{\rm b}(t=0) =
\kappa_{\rm b,0}$ that increases towards $\kappa_{\rm b} (t \to
\infty) = \kappa_{\rm b}^* > \kappa_{\rm b,0} $.  In this case,
negative affects linger less in the minds of former users as recovery
continues.  We mathematically represent the processing rates
$\kappa_{\rm a}(t), \kappa_{\rm b}(t)$ during abstinence as
\begin{subequations}
  \begin{align}
	\kappa_{\rm a}(t) =& \kappa_{\rm a,0} e^{-\gamma_{\rm a} t}
        + \kappa_{\rm a^*} (1 - e^{- \gamma_{\rm a} t}), \label{kappatimeb} \\
	\kappa_{\rm b}(t) =& \kappa_{\rm b,0} e^{-\gamma_{\rm b} t}
        + \kappa_{\rm b^*} (1 - e^{- \gamma_{\rm b} t}),\label{kappatimeb}
  \end{align}
  \end{subequations}
where $\gamma_{\rm a}^{-1}, \gamma_{\rm b}^{-1}$ are typical time
scales associated with neuroadaptive changes to the processing rates
and where $\kappa_{\rm a,0} > \kappa_{\rm a}^*$ and $\kappa_{\rm b,0}
< \kappa_{\rm b}^*$.  The positive affect $M_{\rm a}(t)$ in
Eq.~\ref{Ma_outcome} can thus be written as
\begin{equation}
	M_{\rm a}(t) = \sum_{i,t\ge t_i^{\rm a}}A_i e^{-\kappa^*_{\rm a}(t-t_i^{\rm a})} \exp\left[
	\frac{(\kappa_{\rm a}^*-\kappa_{\rm a, 0})}
	  {\gamma_{\rm a}}\left(e^{-\gamma_{\rm a} t_i^{\rm a}}-e^{-\gamma_{\rm a} t}\right)\right].
        \label{M_a0}
\end{equation}
For $M_{\rm b}(t)$ in Eq.~\ref{Mb_outcome} instead we find
\begin{equation}
  M_{\rm b}(t) = -\sum_{j,t\ge t_j^{\rm b}}B_j e^{-\kappa^*_{\rm b}(t-t_j^{\rm b})}
  \exp\left[
	\frac{(\kappa_{\rm b}^*-\kappa_{\rm b, 0}) }
	     {\gamma_{\rm b}}
             \left(e^{-\gamma_{\rm b} t_j^{\rm b}}-e^{-\gamma_{\rm b} t}\right)\right].\label{M_b0}
\end{equation}
If the restoring, neuroadaptive changes to $\kappa_{\rm a}(t)$ occur
over short time scales such that $\gamma_{\rm a} t_i^{\rm a} \gg 1$,
then $\kappa_{\rm a} (t)$ can be approximated by its equilibration
value $\kappa_{\rm a}^*$.  Conversely, for longer lived changes scales
such that $\gamma_{\rm a} t_i^{\rm a} \ll 1$, then $\kappa_{\rm a}
(t)$ can be approximated by its initial condition $\kappa_{\rm a,0}$.
In either of these two limits, $\kappa_{\rm a}(t)$ can be approximated
by a constant, $\kappa_{\rm a}$. Similar considerations hold for
$\kappa_{\rm b} (t)$ that in the same limits can be modeled as a
constant $\kappa_{\rm b}$. We can thus write
\begin{equation}
	M_{\rm a}(t) =
	\sum_{i, t \geq t_i^{\rm a}}  A_i e^{-\kappa_{\rm a} (t-t_i^{\rm a})}, \quad 
	M_{\rm b}(t) =- \sum_{j, t \geq t_j^{\rm b}}
        B_j e^{-\kappa_{\rm a} (t-t_j^{\rm b})}, \label{Mb_t1app}
\end{equation}
where $\kappa_{\rm a} = \kappa_{\rm a, 0}$ or $\kappa_{\rm a}^*$
depending on the proper limit (and similarly for $\kappa_{\rm b}$) and
use the results for the constant $\kappa_{\rm a}, \kappa_{\rm b}$
cases discussed in Eqs.~\ref{Ma_Mb}.

Instead of considering a sequence of positive or negative events, for
simplicity, we now assume there is a constant negative input
$Y=-0.5$/day and that there are no random events.  In this scenario,
the ideal case of an individual who has never used drugs is
represented by the negative mental state $M(t) = M_{\rm b} (t)$ given
by
\begin{equation}
  M(t) = \frac {Y}{\kappa^*_{\rm b}}(1 - e^{-\kappa_{\rm b}^*t}),
  \label{noexp}
\end{equation}
\noindent
and obtained using the standard processing rate $\kappa^*_{\rm b}$.
We identify the scenario of a recovering addict processing events with
the same rates as if drugs were never used, as a proxy for full recovery.
A patient still in recovery on the other hand processes events at the
time-dependent rate $\kappa_{\rm b} (t)$ given by
Eq.\,\ref{kappatimeb}.  For this individual, the same circumstances
yield the following negative mental state
\begin{equation}
M(t)= Y e^{ - \int_0 ^t \kappa_{\rm b} (t') {\rm d} t' }
\int_0 ^t e^ {\int_0 ^{t'} \kappa_{\rm b} (t'') {\rm d} t''}  \dd t' 
\label{recov}
\end{equation}
Eq.\,\ref{recov} reduces to Eq.\,\ref{noexp} under no recovery,
when $\gamma_{\rm b} = 0$ in Eq.\,\ref{recov}, provided 
$\kappa^*_{\rm b}$ is replaced by $\kappa_{\rm b,0}$ in  Eq.\,\ref{noexp}.

\begin{figure*}[t]
  \centering
\includegraphics[width=0.75\linewidth]{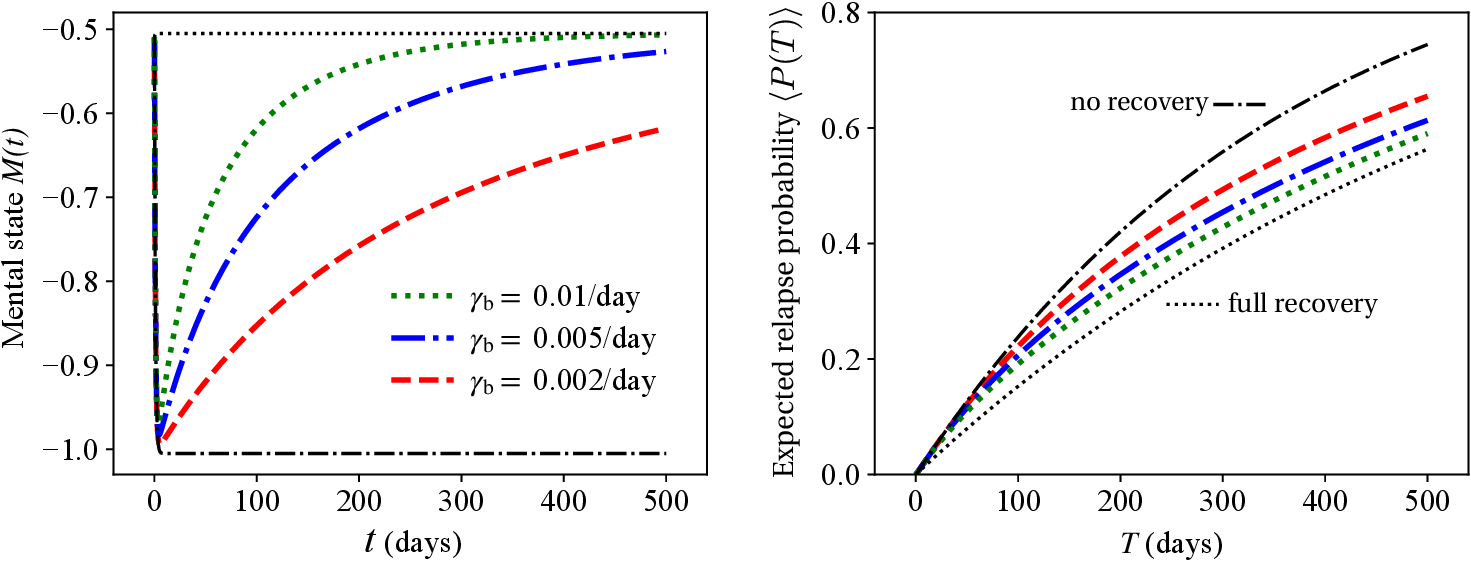}
\caption{Dynamics of $M(t)$ and $\langle P(T) \rangle$ under a
  continuous negative input $Y= - 0.5$/day, an initial processing rate
  $\kappa_{{\rm b}, 0}=1$/day, a fully recovered processing rate
  $\kappa_{\rm b}^*=2$/day, and $R_0=10^{-3}$.  Recovery rates are
  given by $\gamma_\mathrm{b}=0.002, 0.005, 0.01$/day.  The black
  dashed curve is the ideal case of the individual never having used
  drugs, or having fully recovered (Eq.\,\ref{noexp}), and the dotted
  curve is the no recovery case ($\gamma_{\rm b} = 0$ in
  Eq.\,\ref{recov}).  Intermediate curves show patients in recovery
  who tend to approach the ideal case of never having drugs after a
  long enough recovery time.}
	\label{fig:dynamic kappa}
	\end{figure*}

In Fig.\,\ref{fig:dynamic kappa} we consider a dynamically varying
$\kappa_{\rm b}(t)$ and plot the mental state $M(t)$ and relapse
probability $P(t)$ for an initial processing rate $\kappa_{0, \rm b} =
1$/day and the recovered, standard processing rate $\kappa^*_{\rm b} =
2$/day. The initially large value of $\kappa_{0, \rm b}$ implies that
right after cessation of drug use the neurocircuitry of the individual
is still compromised, and any life event is processed quickly.  We
also set $\gamma_{\rm b} = 0.002$/day, $\gamma_{\rm b} = 0.005$/day
and $\gamma_{\rm b} = 0.01$/day, corresponding to recovery times from
$\kappa_{0,\rm b}$ to $\kappa^*_{\rm b}$ ranging from three months to
one and half years, approximately.

The mental states $M(t)$ in each of these scenarios are shown in
Fig.\,\ref{fig:dynamic kappa}(a).  In Fig.\,\ref{fig:dynamic kappa}(b)
we show the corresponding expected relapse probability, $\langle P(T)
\rangle$.  Here, the state of no exposure to drugs (Eq.\,\ref{noexp})
is represented by the lower-bounded curve and the state of no recovery
from drugs (Eq.\,\ref{recov} with $\gamma_{\rm b} = 0$) is represented
by the upper-bounded one.  All other curves correspond to
Eq.\,\ref{recov} with finite, non-zero values of the recovery rate
$\gamma_{\rm b}$.  As can be seen, the latter are all initially closer
to the upper bound, as the recovery effects are minimal at the onset.
However, as the recovery process continues, the
curves start approaching the lower curve corresponding to the ideal
case of the individual never having used drugs in the first place.
Finally, faster recovery processes (larger values of $\gamma_{\rm b}$)
yield lower relapse probabilities.  

  
\section{Estimating the relapse probability}
\label{ou-process}
\noindent
Here, we consider approximations to the expected relapse probability
$\langle P(T) \rangle$ as given by Eq.~\ref{relaprob}

\begin{equation}
	\langle P(T) \rangle = 
	1-\langle S(T) \rangle =  1 - \left \langle
        \exp\left[-R_0 \int_0^T e^{-M(t')} \dd t' \right] \right \rangle.
\label{relaprob2}
\end{equation}
We first expand the exponential in Eq.~\ref{relaprob2} in a
Taylor series
\begin{equation}
 \langle S(T) \rangle = 
 \sum_{n=0}^{\infty} \frac {R_0^n}{n!} (-1)^n \int_0^T \!\!\cdots\!\int_0^T\!
 \langle e^{- M(t^{(1)})}\!\cdots e^{-M(t^{(n)})} \rangle
 \dd t^{(1)}\!\cdots \dd t^{(n)} 
  \label{taylor}
\end{equation}
and note that upon neglecting correlations we can approximate 
the expectation of the products of $e^{- M(t)}$ in
Eq.\,\ref{taylor} as products of expectations so that 
\begin{equation}
    \int_0^T \!\!\cdots\!\int_0^T \langle e^{- M(t^{(1)})}\cdots
    e^{- M(t^{(n)})} \rangle \dd t^{(1)}\!\cdots \dd t^{(n)}
    \approx \left[\int_0^T  \langle e^{- M(t)} \rangle \dd t \right]^n
  \label{taylor2}
  \end{equation}
\noindent
leading to 
\begin{equation}
\label{approx2}
\langle P(T) \rangle \approx 1-\exp
\bigg[- R_0 \int_0^T \langle e^{-M(t)} \rangle \dd t \bigg].
\end{equation}
\noindent
We now evaluate the integral for the $n=1$ summand in Eq.\,\ref{taylor}, 
for which Eq.\,\ref{taylor2} is exact. We find
\begin{equation}
  \begin{aligned}
 \langle e^{- M(t)}\rangle & = 
  \int_{-\infty}^{\infty}\!\!
  e^{-M}  P_{\rm m}(M, t) \,\dd M \\
  \: & = \exp{\left(\frac {\lambda - 2 Y}{ 2 \kappa} \right)} 
  \exp{\left[- \left(M_0 - \frac{Y}{\kappa} \right) 
e^{-\kappa t}  - \frac{\lambda}{2 \kappa}
e^{-2 \kappa t} \right]}.
  \end{aligned}
\label{firstexpand} 
  \end{equation}
We now consider the $n=2$ summand in Eq.\,\ref{taylor} to determine
the conditions under which the approximation in Eq.\,\ref{taylor2}
fails and correlations must be taken into account. Our goal is thus to
evaluate $\langle e^{-M(t^{(1)})} e^{-M(t^{(2)}) }\rangle \equiv
\langle e^{-M(t')} e^{-M(t'')}\rangle$.  To do this, we must consider
the joint probability density $P_{\rm m}(M_1, M_2, t', t'')$ of
finding the mental state $M_1$ at time $t'$ and of finding the mental
state $M_2$ at time $t'' > t'$, conditioned on the previous value
$M_1$ at time $t'$ This is given by
\begin{equation}
P_{\rm m}(M_1, M_2, t', t'') =P_{\rm m}(M_2, t'' | M_1, t') P_{\rm m}(M_1, t')
\end{equation}
\noindent
where $P_{\rm m}(M_1, t')$ is the probability density of finding $M_1$
at $t'$, with the given initial conditions, and where $P_{\rm m}(M_2,
t' | M_1, t')$ is the probability density of finding $M_2$ at $t''$,
conditioned on having $M_1$ at $t'$. The two quantities evolve
according to the Fokker-Planck equation shown in Eq.~\ref{FP} and lead
to
\begin{equation}
  \begin{aligned}
    \langle e^{- M(t')}  & e^{- M(t'')} \rangle \\
    \: & =  \int_{-\infty}^{\infty} 
  \int_{-\infty}^{\infty}\!
  e^{-M_1} e^{-M_2} P_{\rm m}(M_2, M_1, t', t'') \dd M_1 \dd M_2 \\
\:  & = \int_{-\infty}^{\infty} 
\int_{-\infty}^{\infty}\!
e^{-M_1} e^{-M_2} P_{\rm m}(M_2, t'' | M_1, t') P_{\rm m}(M_1, t')
\dd M_1 \dd M_2
  \end{aligned}
  \label{taylor4}
\end{equation}
where
\begin{equation}
  \begin{aligned}
P_{\rm m} (M_1, t') \sim & {\cal N} \left(\mu_1(t'),~\sigma^2(t')\right) \\
P_{\rm m} (M_2, t'' \vert M_1, t') \sim &  {\cal N}
\left(\mu_2(t''- t'),~\sigma^2(t'' - t')\right) 
  \end{aligned}
  \end{equation}
and ${\cal N} (\mu, \sigma)$ is the normal distribution of mean $\mu$
and variance $\sigma$.  The values $\mu_1, \mu_2, \sigma$ are given by
\begin{equation}
  \begin{aligned}
    \mu_1(t') =& \frac{Y}{\kappa} + \left(M_0 - \frac{Y}{\kappa}\right)
    e^{- \kappa t'} \\
\mu_2(t''-t') =& \frac{Y}{\kappa} +
\left(M_1 - \frac{Y}{\kappa}\right) e^{- \kappa (t'' - t')} \\
\sigma^2 (t) =& \frac{\lambda}{\kappa} (1 - e ^{-2 \kappa t}) .
  \end{aligned}
  \end{equation}
We now evaluate Eq.\,\ref{taylor4} to find 
\begin{equation}
  \begin{aligned}
 \langle e^{- M(t')}  e^{- M(t'')} \rangle = & 
 \langle e^{- M(t')} \rangle   \langle e^{- M(t'')} \rangle
 e^{\frac{2 \lambda}{\kappa}
e^{- \kappa t''} \sinh (\kappa t')}
\end{aligned}
  \label{correlateM}
\end{equation}
where $\langle e^{- M(t')} \rangle$ is given in 
Eq.\,\ref{firstexpand}.
\begin{figure*}[t]
  \centering
  \includegraphics[width=0.75\linewidth]{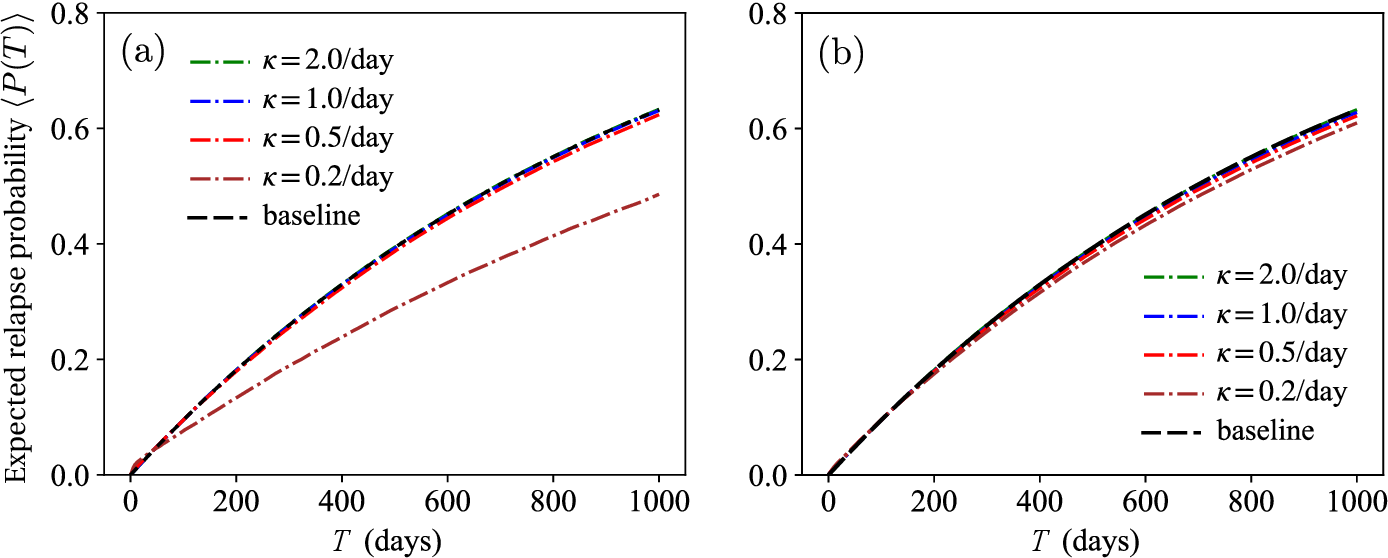}	
\caption{Expected relapse probability $\langle P(T) \rangle$ as
  determined from 5,000 trajectories of the Ornstein-Uhlenbeck process
  with $M_0 =0$ and various parameters $Y, \lambda, \kappa$ and the
  analytical estimate given by Eqs.~\ref{ratenoise} and \ref{approx}.
  The ratio $\lambda = 2 Y$ is kept for all parameter combinations,
  leading to the expectation $\langle P(T) \rangle \approx 1 - \exp
  (R_0 T) $, according to Eq.\,\ref{approx}.  In panel (a) we fix $Y =
  1$/day and $\lambda = 2$/day.  Results from the numerical
  simulations match the analytical estimate only for large values of
  $\kappa$ as illustrated in the text, when correlations can be
  neglected.  In panel (b) we fix $2 \lambda = \alpha \kappa$, with
  $\alpha =8$ to show that under this assumption correlations play a
  less prominent role in the limit $\kappa \to 0$.}
  \label{fig:limits}
\end{figure*}
Given that $t'' > t'$, the exponential term in Eq.\,\ref{correlateM}
will tend towards unity as $\kappa \to \infty$, implying correlations
can be neglected.  The $\kappa \to \infty$ limit corresponds to a
relatively short processing time, consistent with the notion that the
decay of random events is fast and do not allow for strong
correlations. Conversely, if $\kappa \to 0$ we find
\begin{equation}
\lim_{\kappa \to 0} \exp \left(\frac {2 \lambda}{ \kappa}
e^{- \kappa t''} \sinh (\kappa t')  \right) = e^{ 2 \lambda t' }
\end{equation}
which can be quite large for even moderate values of $\lambda$ at
large times. Finally, note that if we set $2 \lambda = \alpha \kappa$
where $\alpha$ is a proportionality constant, then
\begin{equation}
\lim_{\kappa \to 0} \exp \left( \alpha
e^{- \kappa t''} \sinh (\kappa t')  \right) = 1,
\end{equation}
suggesting that an alternative way of neglecting correlations in the
small $\kappa$ limit, when the processing time is large, is to instead
modulate the amplitude of the noise $\lambda$ to be comparable to
$\kappa$.  The evaluation of higher order correlations $n > 2$ in
Eq.\,\ref{taylor} is a more tedious calculation, but we expect that in
the $\kappa \to \infty $ limit the same considerations will apply and
that correlations can also be neglected.  In Fig.\,\ref{fig:limits} we
plot the expected relapse probability $\langle P(T) \rangle$ obtained
by averaging over 5,000 runs of the Ornstein-Uhlenbeck process for
several values of $\kappa$ and other parameters. We show that the
analytical approximation in Eq.\,\ref{approx} holds only for
sufficiently large values of $\kappa$.

\section{Deriving the average relapse rate for Poisson distributed cues}

\noindent
We now derive Eq.\,\ref{mascheroni} assuming cues affect the mental
state $M$ through events of amplitude $w_{\rm peak}$ that are Poisson
distributed with rate $\lambda_{\rm c}$.  According to
Eq.\,\ref{cues}, assuming that within time $t$ there have been $n_{\rm
  c}$ Poisson distributed cues,  the relapse rate is given by
\begin{equation}
	\frac{R(t)}{R_0} =  
	\exp \left[ -\frac{Y}{\kappa_{\rm a}}
          \left(1-e^{-\kappa_{\rm a} t}\right) \right]
	 \!\prod_{\ell =1; t \geq t^{\rm c}_{\ell}}^{n_{\rm c}} 
	 \!\!\!\exp \left[ w_{\rm peak}
           e^{-\kappa_{\rm c} \left(t-t_{\ell}^{\rm c}\right)} \right].
\end{equation}
\noindent
For a general function $f(y)$ one can show that given $n_{\rm c}$
events within time $t$ that are Poisson distributed, the following
holds
\begin{equation}
\label{toshow}
\left. \left \langle \prod_{\ell=1}^{n_{\rm c}} f(t^{\rm c}_{\ell})
\right \rangle \right \vert_{n_{\rm c}}\!\!
=\left[\frac 1 t \int_0^t f(y) \dd y \right]^{n_{\rm c}}
\quad\, n_{\rm c} \sim \mathrm{Poisson}(\lambda_{\rm c},t).
\end{equation}
We will show the validity of this expression below. For now, assuming
Eq.\,\ref{toshow} holds, we write
\begin{equation}
\label{poissoncues}
\left. \left \langle 
\frac{R(t)}{R_0} \right \rangle \right \rvert_{n_{\rm c}}
= \exp\left[-\frac{Y}{\kappa_{\rm a}}
  \left(1-e^{-\kappa_{\rm a} t}\right)\right]
	\left[
	  \frac{1}{t} \int_0^t \exp\left[w_{\rm peak}
            e^{-\kappa_{\rm c} \left(t-t'\right)}\right] 
	\dd t' \right]^{n_{\rm c}}.
\end{equation}
We can now average over the likelihood of having $n_{\rm c}$ events
within $t$ by weighting Eq.\,\ref{poissoncues} by the Poisson
distribution to obtain

\begin{equation}
  \begin{aligned}
\left \langle \frac{R(t)}{R_0} \right \rangle 
= & \exp\left[-\frac{Y}{\kappa_{\rm a}}
  \left(1-e^{-\kappa_{\rm a} t}\right)\right] \\
\: & \times \sum_{n_{\rm c} =0}^{\infty}
\frac{(\lambda_{\rm c} t)^{n_{\rm c}} e^{-\lambda_{\rm c} t}}{n_{\rm c}!}\left[
	  \frac{1}{t} \int_0^t \exp\left[w_{\rm peak}
            e^{-\kappa_{\rm c} \left(t-t'\right)}\right] 
	\dd t' \right] ^{n_{\rm c}}
  \end{aligned}
\end{equation}
Upon evaluating the integral  we can write 
\begin{equation}
  \begin{aligned}
\left \langle \frac{R(t)}{R_0} \right \rangle 
= & \exp\left[ -\lambda_{\rm c} t -\frac{Y}{\kappa_{\rm a}}
  \left(1-e^{-\kappa_{\rm a} t}\right)\right] \\
\: & \times 
\exp\left[ \lambda_{\rm c} \int_0 ^t e^{w_{\rm peak} e^{-\kappa_{\rm c} \left(t-t'\right)}}
  \dd t'  \right]
  \end{aligned}
\end{equation}
which, for  $ \kappa_{\rm a} t, \kappa_{\rm c} t \gg 1$ can be simplified to
\begin{equation}
  \lim_{t \rightarrow \infty}\ln \left\langle\frac{R(t)}{R_0}
  \right\rangle=  
  -\frac{Y}{\kappa_{\rm a}}+\frac{\lambda_{\rm c}}{\kappa_{\rm c}}
  \left({\rm Ei}(w_{\rm peak})-\ln( w_{\rm peak})-\gamma\right)
\end{equation}
where $\gamma$ is the Euler-Mascheroni constant.  To show the validity
of Eq.\,\ref{toshow} we take a general Poisson process of rate $\eta$
and for which $p(t_1, \ldots, t_n | N=n)$ is the probability density
of $n$ events occurring within time $t$ ordered such that $t_1 \leq
t_2 \leq \dots \leq t_n$. In a Poisson process, the possibility of an
event occurring in $[t, t+\dd t]$ is always $\eta \dd t$, which does
not correlate with time, so we can divide the time period $t$ equally
into $M$ segments of length $\dd t$ with $M \gg 1$ so that $t=M \dd
t$.  We label them $ \{[T_1, T_1+\dd t], [T_2, T_2+\dd t], \ldots,
[T_M,T_M+\dd t]\}$. Since each event $t_1 \leq t_k \leq t_n$ will fall
into one of the above segments we can write
\begin{equation}
\begin{aligned}
  p(t_1, \ldots, t_n | & N =n) (\dd t)^n \\
  \: & = \mathbb{P}(T_1<t_1<T_1+\dd t,
  \ldots, T_n<t_n<T_n+\dd t)\\
\: & =\mathbb{P}( \{[T_1, T_1+\dd t], \ldots, [T_n, T_n+\dd t]\}) \\
\end{aligned}
\end{equation}
where the last equality implies that one can simply pick the $n \in M$
segments corresponding to the $t_1 \leq t_k \leq t_n$ events. Since
these intervals are equiprobable, and
\begin{equation}
  p(t_1, \ldots, t_n | N=n) (\dd t)^n
  = \binom{M}{n}^{-1} = \frac{n! (M-n)!}{M!} \approx \frac{n!}{M^n}.
  \nonumber
\end{equation}
Using $M=t/\dd t$, this becomes $p(t_1, \ldots, t_n | N=n)=\frac{n!}{t^n}$.
An alternative way to obtain this is result is to use the explicit
form for the Poisson distribution
\begin{equation}
\begin{aligned}
  p(t_1, \ldots, t_n | N=n) = &
  \frac{p(t_1, \ldots, t_n, N=n)}{P(N=n)} \\
 = & \frac{e^{-\eta (T-t_n)}\cdot\eta e^{-\eta(t_{n-1}-t_n)}\cdots\eta e^{-\eta
		(t_1-t_2)}\cdot\eta e^{-\eta t_{1}}}{\frac{(\eta T)^ne^{-\eta T}}{n!}}\\
= & \frac{\eta^ne^{-\eta T}}{\frac{(\eta T)^ne^{-\eta T}}{n!}}=\frac{n!}{T^n}.
\end{aligned}
\end{equation}
Finally, for a generic function $f(t)$ and for a series of $n$
Poisson-distributed events occurring at times $0 < t_1 < \dots < t_n <
t$ we can write
$$\begin{aligned}
	\left. \left \langle \prod_{j=1}^{n} f(t_j ) \right \rangle \right \vert_n 
	&=\int \left[\prod_{j=1}^{n}f(t_j)
          p(t_1, \ldots, t_n | N=n)\right] \dd t_1\cdots \dd t_n\\
	&=\frac{n!}{t^n}\int \left[\prod_{j=1}^{n}f(t_j)\right]
        \dd t_1 \cdots \dd t_n
        =\frac{n!}{t^n}\int\left[\prod_{j=1}^{n}f(t_j) \dd t_j\right],\\
\end{aligned}$$
where the multidimensional time integrals are constrained by $0 < t_1
< \dots < t_n < t$.  We now eliminate the ordering of the sequence
$t_1,\ldots,t_n$, and divide the integral by the number of
permutations to obtain
$$\begin{aligned}	
\left. \left \langle \prod_{j=1}^{n} f(t_j ) \right \rangle \right \vert_n 
& =\frac{1}{t^n}\left[\prod_{j=1}^{n}\int_{0<t_j<t}f(t_j) \dd t_j\right]\\
\: &=\prod_{j=1}^{n}\left[\int_0^t f(t_j)
  \frac{\dd t_j}{t}\right]=\frac 1 t \left[\int_0^t f(y) \dd y \right]^n, 
\end{aligned}$$
which is Eq.\,\ref{toshow}.


\bibliographystyle{unsrt} 
\bibliography{refs}

\end{document}